\documentclass[onecolumn,12pt,draftcls]{IEEEtran}
\usepackage{amsmath}
\usepackage{amsfonts}
\usepackage{bbding}
\usepackage{amssymb}
\usepackage{array}
\usepackage{subfigure}

\usepackage{graphicx}
\usepackage{subfigure}
\usepackage[named]{algo}
\usepackage{algorithmic}
\usepackage{psfrag}
\usepackage{stfloats}
\usepackage[compress]{cite}
\makeatletter
\renewcommand{\citepunct}{,\penalty\@m\hskip.13emplus.1emminus.1em}
\renewcommand{\citedash}{\hbox{--}\penalty\@m}
\makeatother
\usepackage{setspace}
\usepackage{color}
\allowdisplaybreaks

\usepackage{amsthm}
\usepackage{stfloats}

\newtheorem{pro}{Property}

\usepackage{hyperref}
\usepackage{bm}
\usepackage{bbm}
\renewcommand{\IEEEQED}{\IEEEQEDopen}

\IEEEoverridecommandlockouts

\begin{document}
\title{Improving Network Availability of Ultra-Reliable and Low-Latency Communications with Multi-Connectivity}

\author{
	\IEEEauthorblockN{{Changyang She, Zhengchuan Chen, Chenyang Yang, Tony Q. S. Quek, \\ Yonghui Li, and Branka Vucetic}}
	
	\thanks{This paper was presented in part at the IEEE Vehicular Technology Conference 2017 Spring \cite{Changyang2017TVC}.}

	\thanks{C. She, Y. Li and B. Vucetic are with the School of Electrical and Information Engineering, University of Sydney, Sydney, NSW 2006, Australia (email:shechangyang@gmail.com, \{yonghui.li,branka.vucetic\}@sydney.edu.au).}
		
	\thanks{Z. Chen is with the College of Communication Engineering, Chongqing University, Chongqing 400044, China. Part of this work was done when he was with the Information Systems Technology and Design Pillar, Singapore University of Technology and Design, Singapore 487372 (email: czc@cqu.edu.cn).}
	
	\thanks{C. Yang is with the School of Electronics and Information Engineering, Beihang University, Beijing 100191, China (email:cyyang@buaa.edu.cn).}

	\thanks{T. Q. S. Quek is with the Information Systems Technology and Design Pillar, Singapore University of Technology and Design, 8 Somapah Road, Singapore 487372 (e-mail: tonyquek@sutd.edu.sg). }
}


\maketitle
\begin{abstract}
Ultra-reliable and low-latency communications (URLLC) have stringent requirements on quality-of-service and network availability. Due to path loss and shadowing, it is very challenging to guarantee the stringent requirements of URLLC with satisfactory communication range. In this paper, we first provide a quantitative definition of network availability in the short blocklength regime: the probability that the reliability and latency requirements can be satisfied when the blocklength of channel codes is short. Then, we establish a framework to maximize available range, defined as the maximal communication distance subject to the network availability requirement, by exploiting multi-connectivity. The basic idea is using both device-to-device (D2D) and cellular links to transmit each packet. The practical setup with correlated shadowing between D2D and cellular links is considered. Besides, since processing delay for decoding packets cannot be ignored in URLLC, its impacts on the available range are studied. By comparing the available ranges of different transmission modes, we obtained some useful insights on how to choose transmission modes. Simulation and numerical results validate our analysis, and show that multi-connectivity can improve the available ranges of D2D and cellular links remarkably.
\end{abstract}

\vspace{-0.2cm}
\section{Introduction}
Many mission critical applications require ultra-reliable and low-latency communications (URLLC), such as autonomous vehicles, factory automation, and health care \cite{3GPP2016Scenarios,Popovski2014METIS,Gerhard2014The}. These applications require not only strict end-to-end (E2E) delay (e.g., $1$~ms) and ultra-high reliability (e.g., $10^{-7}$ packet loss probability), but also high network availability (e.g., $99.999$\%) \cite{Popovski2014METIS}.

Network availability is defined as the probability that the QoS (i.e., reliability and latency) of users can be satisfied in a wireless network \cite{Popovski2014METIS}. In the space domain, it is the ratio of
covered area, within which the quality-of-service (QoS) can be satisfied, to the total service area \cite{Mendis2017Availability}. To ensure the stringent QoS requirements of URLLC, the required signal-to-noise ratio (SNR) is high. However, the receive SNR decreases with the communication range, and hence achieving high ratio of covered area to the total service area for URLLC is very challenging.

\subsection{Motivation and Contributions}
To analyze network availability, a semi-analytical signal-to-interference-and-noise ratio (SINR) model was proposed in \cite{David2016WCL}, where the decoding error probability in the short blocklength regime was not considered in the SINR model. Different from traditional video and audio services, achieving ultra-low latency in URLLC requires short blocklength channel codes. It is shown that if Shannon's capacity is used to approximate the achievable rate in the short blocklength regime, the delay and packet loss probability will be underestimated \cite{Gross2015Delay}. In fact, a quantitative definition of network availability in the short blocklength regime is still missing in existing literatures. Further considering that the expression of decoding error probability in the short blocklength regime is very complex \cite{Yury2010Channel}, analyzing network availability in the short blocklength regime is very challenging.

It has been shown that multi-connectivity is an effective way to improve reliability/availability of URLLC \cite{MC2016ICT,MC2016ICCworkshop,Martin2015GCworkshop,Jie2017Availability,Jimmy2017URLLC}. The basic idea is transmitting replicas of each packet over multiple links. If one of the replicas is decoded successfully, then the packet is received. According to the simulation results in \cite{David2016Availability}, the cross-correlation of shadowing of multiple links has significant impacts on network availability. However, analyzing the impacts of the correlation of shadowing on network availability remains an open problem.

Furthermore, the processing delay for decoding packets at the base station (BS) is comparable to the transmission delay in URLLC \cite{3GPPProcessing}. Existing studies implicity assumed that processing delay is much shorter than transmission and queueing delays \cite{Gross2015Delay,She2017CrossLayer,Yulin2016Blocklength}. If processing delay is dominated, it would be better for the BS to amplify-and-forward (AF) the uplink (UL) packets in downlink (DL) transmissions. As a result, how to design transmission schemes when processing delay is considered deserves further study.

Motivated by the above issues, the following questions will be studied in this work: 1) \emph{How to characterize network availability with a mathematic expression that is tractable for analysis?}
2) \emph{How to analyze and optimize network availability in the short blocklength regime with correlated shadowing?} 3) \emph{How to choose different transmission modes, including AF and decode-and-forward (DF) modes when processing delay is considered.}

To address the above challenges, we study how to improve network availability of URLLC, which is equivalent to maximizing the available range defined as the maximal communication distance subject to the network availability requirement. The major contributions of this work are summarized as follows:
\begin{itemize}
\item We establish a framework for analyzing and optimizing available range of different transmission modes, including device-to-device (D2D) mode, cellular modes and multi-connectivity modes that incorporate D2D and cellular links. We provide a quantitative definition of network availability in the short blocklength regime, and derive closed-form expressions of the decoding error probabilities.

\item With the framework, we optimize the transmission durations of different transmission modes to maximize the available range, within which the network availability requirement can be satisfied. The cross-correlation of shadowing between D2D and cellular links is taken into account when optimizing the available range of multi-connectivity modes.

\item We compare available ranges of different transmission modes. Our analysis provides useful insights on how to choose different transmission modes. Simulation and numerical results validate our analysis, and illustrate the impacts of processing delay on the available range.
\end{itemize}

\subsection{Related Work}
There are two lines of related work: applying multi-connectivity for URLLC and analyzing the performance of URLLC in the short blocklength regime.

The studies in \cite{MC2016ICCworkshop} proposed an architecture that uses multiple millimeter wave micro BSs to delivery packets to users, where the control overhead due to mobility was discussed. D2D transmission and retransmission via an access point was studied in \cite{Martin2015GCworkshop} for reducing outage probability in industrial networks. Using coordinated multi-point transmission for improving network availability was considered in \cite{Jie2017Availability}. However, these works did not consider the impact of shadowing, and the reliability was characterized by the outage probability. More recently, path/interface diversity was studied in \cite{Jimmy2017URLLC} to improve the reliability, where each packet is transmitted through multiple paths with different communication interfaces. To study the correlation of failures among multiple paths, a continuous-time-Markov-chain model was used. The studies in \cite{Jimmy2017URLLC} did not focus on radio access network, and the latency of each link was assumed to be Gaussian distributed, but how to achieve the Gaussian distributed latency was not studied.

The achievable rate in the first line of related work is characterized by Shannon capacity, which is the maximal achievable rate when the blocklength of channel codes approaches infinite. To achieve ultra-low latency, the blocklength of channel codes is short in URLLC, and hence the maximal achievable rate in the short blocklength regime should be applied \cite{Yury2010Channel,Yury2014Quasi,Giuseppe2016Toward}.

By using the maximal achievable rate in the short blocklength regime, a cross-layer resource allocation that depends on both channel-state information and queue-state information in DL transmission was optimized in \cite{She2017CrossLayer}. A short packet delivery mechanism was proposed in \cite{She2018Joint}, and the UL and DL resource configurations were jointly optimized. By analyzing the maximal achievable rate in short blocklength regime and effective capacity in relay systems in \cite{Yulin2016Blocklength}, it was found that the DF relaying can achieve higher data rate than direct transmission in the short blocklength regime. However, these works do not take shadowing and processing delay into consideration. Due to the complicated decoding process, decoding at the relay/BS introduces more processing delay than the AF mode. As a result, the DF mode may not be optimal in terms of maximizing network availability.


The rest of this paper is organized as follows. In Section II, we introduce the system model. In Section III, packet loss probabilities with different transmission modes are derived. In Section IV, we establish the framework for improving available ranges of the DF cellular and DF multi-connectivity modes by optimizing durations of different transmission phases. In Section V, we compare the available ranges of different transmission modes. Numerical results are presented in Section VI. Finally, we conclude our work in Section VII.

\section{System Model, QoS Requirements and Network Availability}
\begin{figure}[htbp]
        \vspace{-0.3cm}
        \centering
        \begin{minipage}[t]{0.5\textwidth}
        \includegraphics[width=1\textwidth]{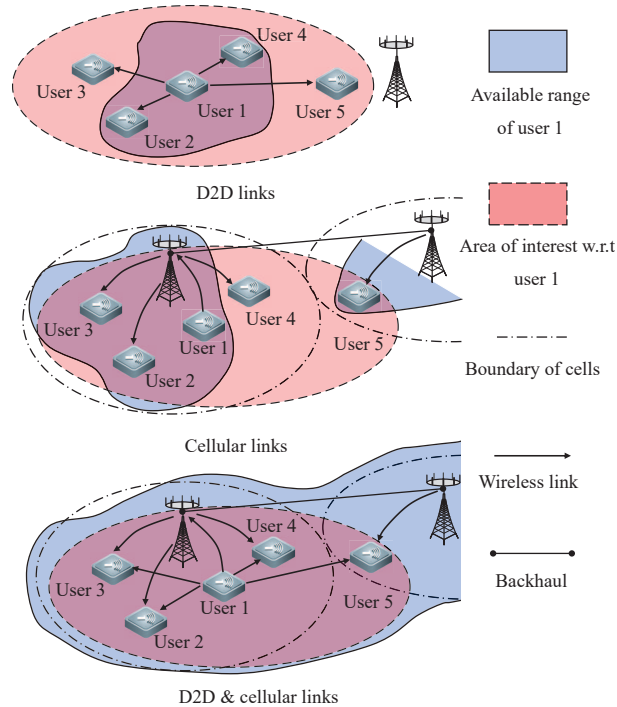}
        \end{minipage}
        \vspace{-0.3cm}
        \caption{System models with different transmission modes.}
        \label{fig:system}
        \vspace{-0.5cm}
\end{figure}

Consider a cellular network, where each BS is equipped with $N_{\rm t}$ antennas, and each single-antenna user either has packets to transmit or requires packets from other users. The users that need to transmit packets are referred to as senders, and the other users are referred to as receivers. As illustrated in Fig. \ref{fig:system}, user $1$ needs to send packets to the users lying in the area of interest with respect to (w.r.t.) it. The range of the area of interest depends on the application scenarios, say a few meters for factory automation, and tens of meters or even longer for autonomous vehicles. If the available range is smaller than the area of w.r.t. user $1$, then the QoS requirement of some target receivers cannot be satisfied.

Three communication scenarios are shown in Fig. \ref{fig:system}. In the first scenario, the packets are transmitted via D2D links, where the BS only participates in coordination and does not transmit packets. In the second scenario, all the packets are sent via cellular links, and D2D transmission is not allowed. The QoS requirement of some users located at the edge of a cell cannot be satisfied. In the third scenario, both D2D and cellular links are used to transmit packets, and hence it is possible to satisfy the QoS requirement of all the users lying in the area of interest w.r.t. user $1$.

\subsection{System Model}
Time is discretized into frames with duration $T_{\rm f}$, which is the minimal time granularity of the system. To ensure ultra-low latency, short frame structure is considered, say $T_{\rm f}$ is around $0.1$~ms \cite{Petteri2015A}. For typical application scenarios in URLLC like machine-type communications and vehicle networks, the packet arrival rate is around $10$ to $30$ packets/s \cite{Hassan2013A,3GPP2012MTC}. For these applications, the inter-arrival time between two packets is longer than $1$~ms, which is the typical E2E delay requirement in URLLC. According to the observations in real use cases, a user generates a packet after the previous packet has been transmitted successfully or discarded due to delay bound violation. Therefore, there is no queue at each user.

\begin{figure}[htbp]
	\vspace{-0.3cm}
	\centering
	\begin{minipage}[t]{0.42\textwidth}
		\includegraphics[width=1\textwidth]{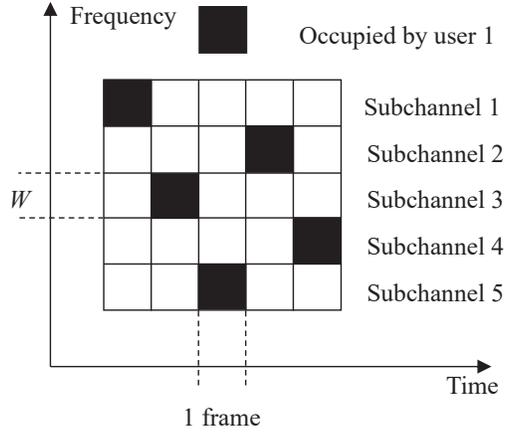}
	\end{minipage}
	\vspace{-0.3cm}
	\caption{Exploiting frequency diversity by frequency-hopping.}
	\label{fig:fd}
	\vspace{-0.5cm}
\end{figure}

In typical scenarios of URLLC, the E2E delay is shorter than the channel coherence time \cite{3GPP2016Scenarios}. If we simply retransmit a packet over successive frames, there will be no diversity gain. However, in practical systems, $W_{\max}$ is larger than the coherence bandwidth. By introducing frequency-hopping in retransmission, each packet can be transmitted over different subchannels in different transmission phases. An example of retransmission scheme is illustrated in Fig. \ref{fig:fd}. In this way, frequency diversity can be exploited to improve reliability.

To avoid interference, orthogonal virtual subchannels are reserved for different senders, and each sender broadcast packets over its reserved virtual subchannel \cite{David2005Fundamentals}. A virtual subchannel is a sequence of subchannels that will be occupied by a sender in the next a few frames. The indices of the subchannels in a virtual subchannel depend on hopping pattern. As illustrated in Fig. \ref{fig:fd}, the virtual subchannel allocated to user $1$ is $(1,3,5,2,4)$, where the $i$th element is the index of the real subchannel that is allocated to sender $1$ in the $i$th frame. When a sender is accessed to one of the BSs, a virtual subchannel is reserved to it. We assume that there are $K$ senders and $K$ virtuals subchannel in the wireless network. Let $W_{\max}$ be the total bandwidth, which is equally allocated to the $K$ senders. As a result, the bandwidth of each virtual subchannel is $W = W_{\max}/K$. How to design hopping pattern to avoid collision can be found in \cite{David2005Fundamentals}, and will not be discussed in this work. With the above bandwidth reservation scheme, there is no scheduling procedure before the transmission of each packet.

\subsection{Transmission Modes}
We consider five transmission modes throughout this paper: D2D mode, AF/DF cellular modes, and AF/DF multi-connectivity modes that consist of D2D and cellular links. For all the transmission modes, we consider a two-phase transmission protocol.

\subsubsection{D2D Mode} With D2D mode, each sender broadcasts its packets to its target receivers in the two phases. Since broadcast is considered, there is no need to estimate channel state information (CSI) at the transmitters.

\subsubsection{AF/DF Cellular Modes} With the cellular modes, one virtual subchannel is used for both UL and DL transmissions. In the first phase, each sender uploads its packet to a BS. In the second phase, the BS simply amplifies and forwards the received signals with AF cellular mode. With DF cellular mode, the BS first decodes the packet, and then broadcasts the packet to receivers if the packet is successfully decoded. With AF cellular mode, the noise in the first phase is amplified in the second phase. With DF cellular mode, processing delay for decoding packets should be considered in URLLC.

\subsubsection{AF/DF Multi-connectivity Modes} AF multi-connectivity mode is a combination of the D2D and AF cellular modes. In the first phase, each sender broadcasts a packet to its target receivers and BSs. We consider the worst case that only the nearest BS can decode the packet from the sender. In this case, if a sender and its target receivers lie in different cells, the packet will be forwarded to farther BSs via backhaul. In the second phase, the sender re-broadcasts the packet and the BSs amplify the signal received in the first phase and broadcast it to the receivers. We consider an intra-network for multi-connectivity modes, where the signals from the sender and the BS are transmitted over the same virtual subchannel in the second phase. We assume that the signals from the sender and the BS are synchronized, and hence the signals contribute to useful signal power \cite{David2017WCNC}. One example of intra-network is the cooperative multi-point \cite{Jie2017Availability}. For each receiver, we consider the worst case that it can only receive the signals from the nearest BS and the D2D link. The receive powers from the other BSs are ignored.

DF multi-connectivity mode is a combination of the D2D and DF cellular modes. The only difference between DF and AF multi-connectivity modes is that the BSs will try to decode the uploaded packets.

{\bf{Remark:}} The broadcast transmission considered in our work is different from the unicast transmission specified in the $5$G New Radio (NR) \cite{3GPP2017NR}. Such a difference results from the communication scenarios. In our work, we consider the scenarios that each sender needs to transmit packets to multiple receivers, which will be common in the future vehicle networks and factory automation \cite{Popovski2014METIS}. For these scenarios, broadcast transmission modes are suitable since there is no overhead caused by control signalling and feedback between the sender and multiple receivers. In the 5G NR, point-to-point communication between a user and a BS is considered. In the point-to-point communication scenario, unicast transmission can achieve better reliability than broadcast, and feedback of CSI and acknowledgement character will not cause very high overhead. Considering that there is a tradeoff between control signalling overhead and QoS, how to optimize control overhead is an important topic for URLLC, and deserves further study.

\subsection{QoS Requirements of URLLC}
The QoS requirements of URLLC can be characterized by the E2E delay of each packet and the packet loss probability, denoted as $D_{\max}$ and ${\varepsilon _{\max }}$, respectively. As discussed in \cite{Changyang2017Mag,Meryem2016Tactile}, possible delay components include transmission delay, queueing delay, and processing delay in radio access network, and backhaul delay.

\begin{figure}[htbp]
	\vspace{-0.2cm}
	\centering
	\begin{minipage}[t]{0.6\textwidth}
		\includegraphics[width=1\textwidth]{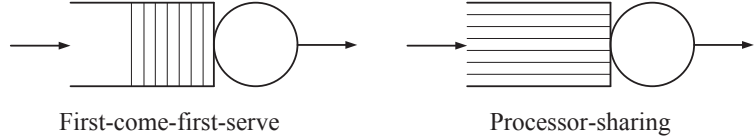}
	\end{minipage}
	\vspace{-0.2cm}
	\caption{Different service order.}
	\label{fig:ps}
	\vspace{-0.4cm}
\end{figure}

If packets are transmitted via BSs and the BSs need to decode the packets, then processing delay for decoding packets should be taken into accout. Although first-come-first-serve (FCFS) server is widely deployed in communication systems, we consider a processor-sharing (PS) server in the computing system of each BS. Both FCFS server and PS server are illustrated in Fig. \ref{fig:ps}. In the $5$G networks, there are both short packets generated by URLLC services and long packets generated by enhanced mobile broadband services. The processing time for decoding a long packet is much longer than the processing time for decoding a short packet. If FCFS server is used, the short packets that arrive at the server after a long packet have to wait for a long time, and the latency requirement cannot be satisfied. In the scenarios with highly dynamic workloads (i.e., the distribution of the required CPU circles to process each packet has a heavy tail), processor sharing server outperforms FCFS server \cite{Mor2013Queue}.

Let $\Omega_{\rm p}$ be the number of CPU cycles required to decode one short packet. The processing rate of each BS is denoted as $\Omega_{\rm b}$ (cycles/frame), which is the CPU cycles per frame. Since the number of short packets in the server does not exceed the number of senders, the precessing delay is bounded by $D_{\rm p} \leq \frac{(K+K_{\rm L})(\Omega_{\rm p} +\Delta)}{\Omega_{\rm b}}T_{\rm f}$, where $K_{\rm L}$ is the number of long packets in the server and $\Delta$ is the overhead caused by the PS server.\footnote{The ideal PS server cannot be practically implemented. The server can be implemented in a time-sharing way that is closed to the PS server, i.e., the service time in each frame is equally allocated to all the packets in the server. Since the server needs to switch among packets, extra overhead should be considered.} We assume the processing time of long packets is much longer than short packets, and hence $K_{\rm L}$ does not change within $D_{\rm p}$.

According to the above discussion, there is no queue at each sender and the BSs. If a sender and its target receiver are respectively connected to two adjacent BSs that are connected with fiber backhaul, then the backhaul delay $D_{\rm b}$ is around $0.1$ ms \cite{Gongzheng2016Backhaul}, and does not exceed the frame duration in our work, i.e., $D_{\rm b} \leq T_{\rm f}$.

\begin{figure}[htbp]
	\vspace{-0.2cm}
	\centering
	\begin{minipage}[t]{0.5\textwidth}
		\includegraphics[width=1\textwidth]{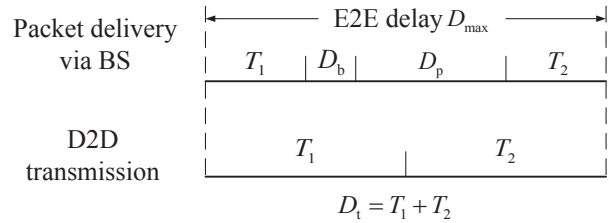}
	\end{minipage}
	\vspace{-0.2cm}
	\caption{Components of E2E delay.}
	\label{fig:E2Edelay}
	\vspace{-0.6cm}
\end{figure}

The delay components are illustrated in Fig. \ref{fig:E2Edelay}, where the transmission delay of a packet $D_{\rm t}$ is divided into two phases with duration $T_1$ and $T_2$, respectively. The E2E delay requirement can be satisfied with the following constraint,
\begin{align}
D_{\rm p} + D_{\rm b} + D_{\rm t} \leq D_{\max}.\label{eq:E2E}
\end{align}
For D2D mode $D_{\rm p} = D_{\rm b} = 0$.

Let $\mu_{\rm sb}$, $\mu_{\rm br}$, and $\mu_{\rm sr}$ be the large-scale channel gains from a sender to the BS (i.e., UL), from the BS to the receiver (i.e., DL) and from the sender to a target receiver (i.e., D2D link), respectively. Given the large-scale channel gains, the packet loss probability can be expressed as $\Pr\{\mathcal{L}|\mu_{\rm sb},\mu_{\rm br},\mu_{\rm sr}\}$, where $\mathcal{L}$ represents the event that a packet is lost. Then, the requirement on packet loss probability can be expressed as follows,
\begin{align}
\Pr\{\mathcal{L}|\mu_{\rm sb},\mu_{\rm br},\mu_{\rm sr}\} \leq {\varepsilon _{\max }}.\label{eq:relia}
\end{align}
$\Pr\{\mathcal{L}|\mu_{\rm sb},\mu_{\rm br},\mu_{\rm sr}\}$ may be independent of $\mu_{\rm sb}$, $\mu_{\rm br}$, or $\mu_{\rm sr}$, but \eqref{eq:relia} can be used to characterize the reliability requirement with different modes. For example, in the cellular modes $\Pr\{\mathcal{L}|\mu_{\rm sb},\mu_{\rm br},\mu_{\rm sr}\}$ is independent of $\mu_{\rm sr}$, we only need to substitute  $\Pr\{\mathcal{L}|\mu_{\rm sb},\mu_{\rm br},\mu_{\rm sr}\} = \Pr\{\mathcal{L}|\mu_{\rm sb},\mu_{\rm br}\}$ into \eqref{eq:relia}.

\subsection{Network Availability and Available Range}
To characterize the network availability, we first relate the large-scale channel gains with the communication ranges as follows: \cite{WirelessCom}
\begin{align}
&10\log_{10}\mu_{\rm sb}\; \text{(dB)} = -10{\alpha}\log_{10}(r_{\rm sb}) \; \text{(dB)} + \delta_{\rm sb} \; \text{(dB)} +\mu_0 \; \text{(dB)}\label{eq:UL},\\
&10\log_{10}\mu_{\rm br} \text{(dB)} = -10{\alpha}\log_{10}(r_{\rm br})\; \text{(dB)} + \delta_{\rm br}\; \text{(dB)}+\mu_0\; \text{(dB)}\label{eq:DL},\\
&10\log_{10}\mu_{\rm sr} \text{(dB)} = -10{\alpha}\log_{10}(r_{\rm sr})\; \text{(dB)} + \delta_{\rm sr}\; \text{(dB)}+\mu_0\; \text{(dB)}\label{eq:DD},
\end{align}
where $r_{\rm sb}$, $r_{\rm br}$, and $r_{\rm sr}$ are the distances between sender and its associated BS, BS and target receiver, and sender and its target receiver, respectively, $\delta_{\rm sb}$,  $\delta_{\rm br}$, and $\delta_{\rm sr}$ are the shadowing of UL, DL and D2D link, respectively, $\alpha > 0$ is the path-loss exponent, and $\mu_0$ is the large-scale channel gain when the communication distance is $1$~m and it depends on antenna characteristics.

The shadowing follows a lognormal distribution with zero mean and $\sigma$~dB standard deviation \cite{3GPPQoS}. Therefore, \eqref{eq:relia} cannot be satisfied with probability one. The network availability is defined as the probability $P_{\rm A}$ that the QoS of each user can be satisfied in a wireless network \cite{Popovski2014METIS}. Given the QoS requirement in \eqref{eq:E2E} and \eqref{eq:relia}, the network availability can be satisfied if the following inequality holds
\begin{align}
\Pr\{D_{\rm p} + D_{\rm b} + D_{\rm t} \leq D_{\max}, \Pr\{\mathcal{L}|\mu_{\rm sb},\mu_{\rm br},\mu_{\rm sr}\} \leq {\varepsilon _{\max }}\} \geq P_{\rm A}.\label{eq:na}
\end{align}
\eqref{eq:na} means that the E2E delay and packet loss probability can be satisfied with probability $P_{\rm A}$.

The communication ranges of cellular and D2D links with guaranteed QoS and availability could be small due to path loss and shadowing. The available range between users and BS is denoted as $r^{\rm c}_{\rm A}$, which is the distance that \eqref{eq:na} can be satisfied for arbitrary $r_{\rm sb} \leq r^{\rm c}_{\rm A}$ and $r_{\rm br} \leq r^{\rm c}_{\rm A}$. Similarly, for D2D transmission, the available range $r^{\rm d}_{\rm A}$ is the distance that \eqref{eq:na} can be satisfied for arbitrary $r_{\rm sr} \leq r^{\rm d}_{\rm A}$.

\section{Packet Loss Probabilities of Different Transmission Modes}
In this section, we derive the packet loss probabilities of different transmission modes, which is useful for analyzing network availability. Considering that the difference between single-cell and multi-cell scenarios lies in the backhaul delay, we only consider a single-cell scenario for notational simplicity.

\subsection{D2D Mode}
Without CSI at the senders, each sender simply broadcasts it packets with the maximal transmit power in order to maximize the available range. By sending pilots, CSI is available at the receivers. With short transmission duration, the blocklength of channel codes is short. According to \cite{Yury2014Quasi}, the achievable rate in short blocklength regime can be accurately approximated as follows,
\begin{align}
R \approx \frac{{W}}{{\ln 2}}\left[ {\ln \left( {1 + \frac{{{\mu _{\rm{sr}}}{P^{\rm t}_{\rm{s}}}{g_{\rm{sr},i}}}}{{{N_0}W}}} \right) - \sqrt {\frac{V}{{{T_i}W}}} f_{\rm{Q}}^{ - 1}\left( {{\varepsilon^{\rm d}_{\rm{sr},i}}} \right)} \right]\; (\text{bits/s}), i=1,2,\label{eq:R}
\end{align}
where $P^{\rm t}_{\rm s}$ is the maximal transmit power of a sender, $N_0$ is the single-side noise spectral density, $g_{\rm{sr},i}$ is the  small-scale channel gain in the $i$th phase, $ f_{\rm{Q}}^{-1}(.)$ is the inverse of Q-function, $\varepsilon^{\rm d}_{\rm{sr},i}$ is the decoding error probability in the $i$th phase, and $V = 1 - {1}/{{{{\left[ {1 + {{{\mu _{\rm{sr}}}{g_{\rm{sr},i}}{P^{\rm t}_{\rm{s}}}}}/{({{N_0}W})}} \right]}^2}}}$. The blocklength of channel codes is $m_{\rm{b}} = T_iW$, which is the number of symbols in the block.

According to \cite{Yury2014Quasi}, the decoding error probability when transmitting $b$~bits from a sender to a receiver in the short blocklength regime can be obtained from \eqref{eq:R} by setting ${T_i}R = b$, i.e.,
\begin{align}
{\varepsilon^{\rm d}_{\rm{sr},i}} \approx {{\mathbb{E}}_{{{g}}_{{\rm sr},i}}}\left\{ {{f_{\rm Q}}\left( {\sqrt {\frac{{{T_i}W}}{V}} \left[ {\ln \left( {1 + \frac{{{\mu _{\rm{sr}}}{g_{\rm{sr},i}}{P^{\rm t}_{\rm{s}}}}}{{{N_0}W}}} \right) - \frac{{b\ln 2}}{{{T_i}W}}} \right]} \right)} \Bigg|\mu _{\rm{sr}}\right\},i=1,2,\label{eq:eu}
\end{align}
where the average is taken over small-scale channel gain conditioned on large-scale channel gain. Given large-scale channel gain, the decoding error probabilities in the two phases only depend on small-scale channel fading. Then, the packet loss probability under the D2D mode can be expressed as follows,
\begin{align}
\Pr\{\mathcal{L}^{\rm d}|\mu _{\rm{sr}}\} \approx &{{\mathbb{E}}_{{{g}}_{{\rm sr},1},{{g}}_{{\rm sr},2}}}\Bigg\{ {{f_{\rm Q}}\left( {\sqrt {\frac{{{T_1}W}}{V}} \left[ {\ln \left( {1 + \frac{{{\mu _{\rm{sr}}}{g_{\rm{sr},1}}{P^{\rm t}_{\rm{s}}}}}{{{N_0}W}}} \right) - \frac{{b\ln 2}}{{{T_1}W}}} \right]} \right)} \times \nonumber\\
&{{f_{\rm Q}}\left( {\sqrt {\frac{{{T_2}W}}{V}} \left[ {\ln \left( {1 + \frac{{{\mu _{\rm{sr}}}{g_{\rm{sr},2}}{P^{\rm t}_{\rm{s}}}}}{{{N_0}W}}} \right) - \frac{{b\ln 2}}{{{T_2}W}}} \right]} \right)}
\Bigg|\mu _{\rm{sr}}\Bigg\} \nonumber\\
\buildrel\textstyle{_{(\text{a})}}\over=&{\varepsilon^{\rm d}_{\rm{sr},1}}{\varepsilon^{\rm d}_{\rm{sr},2}},\label{eq:D2D}
\end{align}
where $\left(\text{a}\right)$ holds when ${{g}}_{{\rm sr},1}$ and ${{g}}_{{\rm sr},2}$ are independent, which is the case with frequency-hopping.

\subsection{Cellular Modes}

\subsubsection{AF Cellular Mode} With AF cellular mode, the processing delay is zero, i.e., $D_p = 0$. However, the noise in the first phase is amplified in the second phase. Denote the received SNR in the first phase at the BS as
\begin{align}
\beta \triangleq \frac{{{\mu _{\rm{sb}}}{g_{\rm{sb}}}{P^{\rm t}_{\rm{s}}}}}{{{N_0}W}}, \label{eq:beta}
\end{align}
where $g_{\rm sb}$ is the UL small-scale channel gain.  When the BS broadcasts a packet with transmit power $P_{\rm b}^{\rm t}$, the signal and noise are amplified with transmit power $ {\beta P_{\rm b}^{\rm t}}/{(\beta+1)} $ and $ {P_{\rm b}^{\rm t}}/{(\beta+1)} $, respectively. In DL transmission, the BS does not have CSI. Then, the received signal power is ${\beta\mu_{\rm br}g_{\rm br}P_{\rm b}^{\rm t}}/{[(\beta+1)N_{\rm t}]}$, and the noise power is ${\mu_{\rm br}g_{\rm br}P_{\rm b}^{\rm t}}/{[(\beta+1)N_{\rm t}]}+N_0W$. Therefore, the final SNR can be expressed as
\begin{align}\label{eq:SNRC}
\gamma^{\rm c,I} = \frac{{\beta\mu_{\rm br}g_{\rm br}P_{\rm b}^{\rm t}}}{{\mu_{\rm br}g_{\rm br}P_{\rm b}^{\rm t}}+{(\beta+1)N_{\rm t}}N_0W}.
\end{align}

To forward the same symbols with the same bandwidth in two phases, the transmission durations in the two phases should be the same, i.e., $T_1 = T_2 = D_{\rm t}/2$.\footnote{The relation between the number of symbols in each block and time/frequency resources is $m_{\rm b} = WT_i$, $i=1,2$.} Similar to \eqref{eq:eu}, the packet loss probability (i.e., decoding error probability) with the AF cellular mode is
\begin{align}\label{eq:CI}
\Pr\{\mathcal{L}^{\rm c,I}|\mu_{\rm sb},\mu_{\rm br}\} \approx {{\mathbb{E}}_{{{g}}_{{\rm sb}},g_{\rm br}}}\left\{ {{f_{\rm Q}}\left( {\sqrt {\frac{{{D_{\rm t}}W}}{2V}} \left[ {\ln \left( {1 + \gamma^{\rm c,I} } \right) - \frac{{2b\ln 2}}{{{D_{\rm t}}W}}} \right]} \right)} {\Bigg{|}}\mu_{\rm{sb}},\mu_{\rm{br}}\right\}.
\end{align}

\subsubsection{DF Cellular Mode} Denote the decoding error probabilities in the UL and DL  phases with the DF cellular mode as ${\varepsilon^{\rm c, II} _{\rm{sb}}}$ and ${\varepsilon^{\rm c, II} _{\rm{br}}}$, respectively. Similar to \eqref{eq:eu}, ${\varepsilon^{\rm c, II} _{\rm{sb}}}$ can be derived as follows,
\begin{align}\label{eq:eUL}
{\varepsilon^{\rm c, II} _{\rm{sb}}} \approx {{\mathbb{E}}_{{{g}}_{{\rm sb}}}}\left\{ {{f_{\rm Q}}\left( {\sqrt {\frac{{{T_1}W}}{V}} \left[ {\ln \left( {1 + \frac{{{\mu_{\rm{sb}}}{g_{\rm{sb}}}{P^{\rm t}_{\rm{s}}}}}{{{N_0}W}}} \right) - \frac{{b\ln 2}}{{{T_1}W}}} \right]} \right)} {\Bigg{|}}\mu _{\rm{sb}}\right\}.
\end{align}
${\varepsilon^{\rm c, II} _{\rm{br}}}$ can be obtained from \eqref{eq:eUL} by substituting $P_{\rm s}^{\rm t}$, $g_{\rm sb}$, $\mu _{\rm{sb}}$, and $T_{\rm 1}$ with $P_{\rm b}^{\rm t}/N_{\rm t}$, $g_{\rm br}$, $\mu _{\rm{br}}$, and $T_{\rm 2}$, respectively, where $g_{\rm br}$ is the small-scale channel gain from the BS to the receiver. A packet is lost when either UL or DL transmission fails. Therefore, the packet loss probability with the DF cellular mode is
\begin{align}
\Pr\{\mathcal{L}^{\rm c,II}|\mu_{\rm sb},\mu_{\rm br}\} & = 1-(1-\varepsilon^{\rm c, II} _{\rm{sb}})(1-{\varepsilon^{\rm c, II} _{\rm{br}}}) = {\varepsilon^{\rm c, II} _{\rm{sb}}} + {\varepsilon^{\rm c, II} _{\rm{br}}} - {\varepsilon^{\rm c, II} _{\rm{sb}}}  {\varepsilon^{\rm c, II} _{\rm{br}}}.\label{eq:CII}
\end{align}

\subsection{Multi-connectivity Modes}

\subsubsection{AF Multi-connectivity Mode} Similar to the AF cellular mode, to transmit the same symbols with the same bandwidth in the two phases, the transmission durations of the two phases should be the same, $T_1 = T_2 = D_{\rm t}/2$. The decoding error probability in the first phase ${\varepsilon^{\rm m,I}_{1}} $ is the same as ${\varepsilon^{\rm d}_{\rm{sr},1}} $ in \eqref{eq:eu} with $T_1 = D_{\rm t}/2$. To derive the decoding error probability in the second phase, we need to derive the receive SNR first. The sender broadcasts the packet with transmit power $P^{\rm t}_{\rm{s}}$, and the BS broadcasts signal and noise with transmit power $ \frac{\beta}{\beta+1} P_{\rm b}^{\rm t}$ and $ \frac{1}{\beta+1} P_{\rm b}^{\rm t}$, respectively. Thus, the received signal power is $\frac{\beta\mu_{\rm br}g_{\rm br}P_{\rm b}^{\rm t}}{(\beta+1)N_{\rm t}}  + \mu_{\rm sr}g_{\rm sr}P^{\rm t}_{\rm{s}}$ and the total noise power is $\frac{\mu_{\rm br}g_{\rm br}P_{\rm b}^{\rm t}}{(\beta+1)N_{\rm t}}+N_0W$. Then, the received SNR is given by
\begin{align}\label{eq:SNRHI}
\gamma^{\rm m,I} = \frac{{\beta\mu_{\rm br}g_{\rm br}P_{\rm b}^{\rm t}}  + (\beta+1)N_{\rm t}\mu_{\rm sr}g_{\rm sr}P^{\rm t}_{\rm{s}}}{{\mu_{\rm br}g_{\rm br}P_{\rm b}^{\rm t}}+(\beta+1)N_{\rm t}N_0W}.
\end{align}

Given the receive SNR, the decoding error probability in the second phase is
\begin{align}\label{eq:decodeM1}
{\varepsilon^{\rm m,I}_2} \approx {{\mathbb{E}}_{{{g}}_{{\rm sb}},g_{\rm br},g_{\rm sr}}}\left\{ {{f_{\rm Q}}\left( {\sqrt {\frac{{{D_{\rm t}}W}}{2V}} \left[ {\ln \left( {1 + \gamma^{\rm m, I} } \right) - \frac{{2b\ln 2}}{{{D_{\rm t}}W}}} \right]} \right)} {\Bigg{|}}\mu_{\rm{sb}},\mu_{\rm{br}},\mu _{\rm{sr}}\right\}.
\end{align}
The packet is lost when the transmissions in both phases fail. Thus, the packet loss probability can be expressed as follows,
\begin{align}\label{eq:HI}
\Pr\{\mathcal{L}^{\rm m,I}|\mu_{\rm{sb}},\mu_{\rm{br}},\mu _{\rm{sr}}\} = {\varepsilon^{\rm m,I}_1}{\varepsilon^{\rm m,I}_2}.
\end{align}

\subsubsection{DF Multi-connectivity Mode} In the first phase, the decoding error probabilities at the receiver and the BS are denoted as $ \varepsilon^{\rm d}_{{\rm sr,}1}$ and ${\varepsilon^{\rm c, II} _{\rm{sb}}}$, respectively. If the packet is not successfully decoded at the BS, then the decoding error probability in the second phase is equal to $ \varepsilon^{\rm d}_{{\rm sr,}2}$. Otherwise, the decoding error probability is
\begin{align}
{\varepsilon_2^{\rm{m,II}}} \approx {{\mathbb{E}}_{g_{\rm{br}},g_{\rm{sr}}}}\left\{ {f_{\rm{Q}}}\left( {\sqrt { \frac{{T_2}W}{V} } }\left[{\ln \left( 1 +  \gamma^{\rm m,II} \right) - \frac{{b\ln 2}}{{{T_2}W}}}\right] \right){\Bigg{|}}\mu_{\rm br},\mu_{\rm sr} \right\},\label{eq:eDL}
\end{align}
where
\begin{align}\label{eq:SNRH2}
\gamma^{\rm m,II} = \frac{{{\mu_{\rm{br}}}{g_{\rm{br}}}{P^{\rm t}_{\rm{b}}}/{N_{\rm{t}}}+{\mu _{\rm{sr}}}{g_{\rm{sr}}}{P^{\rm t}_{\rm{s}}}}}{{{N_0}W}}.
\end{align}

For the DF multi-connectivity mode, a packet is lost in two cases. If the D2D transmission in the first phase fails but the packet is successfully received at the BS, then the packet will be lost if it is not successfully decoded by the target receiver in the second phase. The probability that this case happens is $\varepsilon^{\rm d}_{{\rm sr,}1}(1 - {\varepsilon^{\rm c, II} _{\rm{sb}}}){\varepsilon_2^{\rm{m,II}}}$. If both the D2D and UL transmissions fail in the first phase, then the packet will be lost if the D2D transmission in the second phase also fails. The probability that this case happens is $\varepsilon^{\rm d}_{{\rm sr,}1} {\varepsilon^{\rm c, II} _{\rm{sb}}}\varepsilon^{\rm d}_{{\rm sr,}2}$. Since the two cases are mutual exclusive, the packet loss probability of the DF multi-connectivity mode can be expressed as follows,
\begin{align}
\Pr\{\mathcal{L}^{\rm m,II}|\mu_{\rm{sb}},\mu_{\rm{br}},\mu _{\rm{sr}}\} = \varepsilon^{\rm d}_{{\rm sr,}1}(1 - {\varepsilon^{\rm c, II} _{\rm{sb}}}){\varepsilon_2^{\rm{m,II}}} + \varepsilon^{\rm d}_{{\rm sr,}1} {\varepsilon^{\rm c, II} _{\rm{sb}}}\varepsilon^{\rm d}_{{\rm sr,}2}.\label{eq:loss}
\end{align}

\subsection{Method for Deriving Decoding Error Probabilities}
To obtain the packet loss probabilities with different transmission modes, we need first derive the decoding error probabilities in \eqref{eq:eu}, \eqref{eq:CI}, \eqref{eq:eUL}, \eqref{eq:decodeM1}, and \eqref{eq:eDL}. Since there is no closed-form expression of Q-function, it is very challenging to derive the decoding error probabilities. To overcome this difficulty, we introduce an approximation of ${f_{\rm Q}}\left( {\frac{{{{\log }_2}\left( {1 + \gamma } \right) - {r_{\rm{c}}}}}{{\sqrt {V\left( \gamma  \right){{\left( {{{\log }_2}e} \right)}^2}/{m_{\rm{b}}}} }}} \right)\approx \Lambda \left( \gamma  \right)$ \cite{Behrooz2014WCL}, where $\gamma$ is the receive SNR, $r_{\rm c} = b/m_{\rm b}$ is the number of bits in each symbol,
\begin{align}\label{eq:Taylor}
\Lambda \left( \gamma  \right) \triangleq \left\{ {\begin{array}{*{20}{c}}
{1,{{\;\;\;\;\;\;\;\;\;\;\;\;\;\;\;\;\;\;\;\;\;\;\;\;\;\;\;\;\;\;\;\;\;}}\gamma  \le \zeta }\\
{\frac{1}{2} - \omega \sqrt {{m_{\rm{b}}}} \left( {\gamma  - \theta } \right),\zeta  < \gamma  < \xi }\\
{0,{\;\;\;\;\;\;\;\;\;\;\;\;\;\;\;\;\;\;\;\;\;\;\;\;\;\;\;\;\;\;\;\;\;}\gamma  \ge \xi }
\end{array}} \right.,
\end{align}
where $\omega = \frac{1}{2\pi\sqrt{2^{2r_{\rm c}}-1}}$, $\theta = 2^{r_{\rm c}}-1$, $\zeta = \theta-\frac{1}{2\omega\sqrt{m_{\rm b}}}$, and $\xi = \theta+\frac{1}{2\omega\sqrt{m_{\rm b}}}$. With the above approximation, the decoding error probability can be expressed as \cite{Gu2018WCL}
\begin{align}
{\mathbb{E}}\left\{ {{f_Q}\left( {\frac{{{{\log }_2}\left( {1 + \gamma } \right) - {r_{\rm{c}}}}}{{\sqrt {V\left( \gamma  \right){{\left( {{{\log }_2}e} \right)}^2}/{m_{\rm{b}}}} }}} \right)} \right\} \approx \omega \sqrt {{m_{\rm{b}}}} \int_\zeta ^\xi  {{F_\gamma }\left( x \right){\rm d}x}, \label{eq:DE}
\end{align}
where $F_{\gamma}(x) = \Pr\{\gamma \leq x\}$ is the cumulative probability function (CDF) of SNR. We will illustrate the impacts of the approximation on network availability via simulation.

From \eqref{eq:DE}, we can derive the decoding error probability over single-input-multiple-output (SIMO) Rayleigh fading channel, i.e., the decoding error probability in the uplink phase of the DF cellular mode (see proof in Appendix \ref{App:SIMO}),
\begin{align}\label{eq:SIMO}
{\varepsilon^{\rm c, II} _{\rm{sb}}} \approx  \frac{\omega \sqrt {{m_{\rm{b}}}} \mu_{\rm sb}P^{\rm t}_{\rm s}}{N_0W}\left[(g_{\rm U}-g_{\rm L})-\sum_{n=0}^{N_{\rm t}}(N_{\rm t}-n)A_n\right],
\end{align}
where $g_{\rm U} = \frac{N_0W\xi}{\mu_{\rm sb}P^{\rm t}_{\rm s}}$, $g_{\rm L} = \frac{N_0W\zeta}{\mu_{\rm sb}P^{\rm t}_{\rm s}}$, $A_n = \frac{g_{\rm L}^n}{n!} e^{-g_{\rm L}} - \frac{g_{\rm U}^n}{n!}e^{-g_{\rm U}}$, and the probability density function (PDF) of $g_{\rm sb}$ over SIMO Rayleigh fading channel is $f_g(x) = \frac{1}{(N_{\rm t}-1)!}x^{N_{\rm t}-1}e^{-x}$. For multiple-input-single-output systems, the decoding error probability is similar to \eqref{eq:SIMO}.

With the above results, we can obtain the packet loss probability of D2D and DF cellular modes. For the AF cellular modes, the CDF of \eqref{eq:SNRC} can be found in \cite{Samy2012Exact}, and will not be discussed in this paper. For DF multi-connectivity mode, the closed-form expression of the CDF of $\gamma^{\rm m,II}$ can be derived as follows (see proof in Appendix \ref{App:CDF}.),
\begin{align}\label{eq:CDFmII}
F_{\gamma^{\rm m, II}}(x) = -\exp(-\frac{x}{C_{\rm sr}})B_1(x)+B_2(x),
\end{align}
where $B_1(x)=\left\{1-\exp\left[\frac{x(C_{\rm sr}-C_{\rm br})}{C_{\rm sr}C_{\rm br}}\right]\sum_{n=0}^{N_{\rm t}-1}{\frac{\left[\frac{x(C_{\rm sr}-C_{\rm br})}{C_{\rm sr}C_{\rm br}}\right]^n}{n!}}\right\}\left(\frac{C_{\rm sr}}{C_{\rm sr}-C_{\rm br}}\right)^{N_{\rm t}}$, $B_2(x) = 1-\exp(-\frac{x}{C_{\rm br}}) \times\sum_{n=0}^{N_{\rm t}-1}{\frac{(\frac{x}{C_{\rm br}})^n}{n!}}$, $C_{\rm sr} = \frac{\mu_{\rm br}P^{\rm t}_{\rm s}}{N_0W}$, and $C_{\rm br} = \frac{\mu_{\rm br}P^{\rm t}_{\rm b}}{N_0WN_{\rm t}}$. By substituting \eqref{eq:CDFmII} into \eqref{eq:DE}, we can obtain \eqref{eq:eDL}. Unfortunately, there is no closed-form expression of the CDF of $\gamma^{\rm m,I}$. To obtain the decoding error probability of the AF multi-connectivity mode, we need to compute the triple integrals in \eqref{eq:decodeM1} numerically.

\section{Framework for Improving Available Ranges of D2D and DF Modes}
With the AF cellular and AF multi-connectivity modes, the same number of symbols are transmitted in the two phases. Given the same bandwidth in the two phases, the transmission durations are equal. Therefore, we cannot adjust transmission durations with the AF modes. However, it is possible to adjust the transmission durations of the two phases with DF modes. For example, if the BS uses different modulation and coding schemes in the two phases, the numbers of symbols that are used to convey the same number of bits are different, and hence $T_1$ and $T_2$ can be different.


In this section, we fixed the total transmission delay as $D_{\rm t} = D_{\max}-D_{\rm p}-D_{\rm b}$, and study how to maximize the available ranges of the DF modes by optimizing the transmission durations in the two phases for a given transmission delay $D_{\rm t}$. For given delay components that satisfy the E2E delay requirement, the network availability requirement in \eqref{eq:na} can be simplified as follows,
\begin{align}\label{eq:sna}
\Pr\{ \Pr\{\mathcal{L}|\mu_{\rm sb},\mu_{\rm br},\mu_{\rm sr}\} \leq {\varepsilon _{\max }}\} \geq P_{\rm A}.
\end{align}

Since large-scale channel gains decrease with the communication distance, given the distribution of shadowing, if \eqref{eq:sna} can be satisfied with $r_{\rm sb}=r_{\rm br}=r_{\rm A}^{\rm c}$, then it can also be satisfied for arbitrary $r_{\rm sb},r_{\rm br} \leq r_{\rm A}^{\rm c}$. Therefore, we consider the worst case that $r_{\rm sb}=r_{\rm br}=r_{\rm A}^{\rm c}$. Similarly, $r_{\rm sr}=r_{\rm A}^{\rm d}$ is considered in the following analysis.

Define ${{\mathbbm{1}}}({\mathcal{A}})$ as an indicator function with respect to event ${\mathcal{A}}$. If event ${\mathcal{A}}$ is true, then ${\mathbbm{1}}({\mathcal{A}}) = 1$. Otherwise, ${\mathbbm{1}}({\mathcal{A}}) = 0$. According to the definition, it is straightforward to see $\Pr\{{\mathcal{A}}\} = {\mathbb{E}}[{\mathbbm{1}}({\mathcal{A}})]$.

\begin{figure}[htbp]
	\vspace{-0.2cm}
	\centering
	\begin{minipage}[t]{0.35\textwidth}
		\includegraphics[width=1\textwidth]{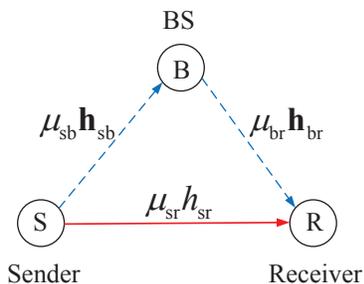}
	\end{minipage}
	\vspace{-0.2cm}
	\caption{Communication distances and shadowing values.}
	\label{fig:shadowing}
	\vspace{-0.8cm}
\end{figure}

\subsection{DF Cellular Mode}

In the DF cellular mode, the cross-correlation of shadowing is taken into consideration. The communication distances and shadowing values of all the links are illustrated in Fig. \ref{fig:shadowing}. A simple model that is widely used to characterize cross-correlation of shadowing is the absolute-distance only model \cite{Thompson2010Shadowing}. For example, the correlation coefficient of shadowing of the cellular links in Fig. \ref{fig:shadowing} can be expressed as
\begin{align}\label{eq:cor}
\rho(\delta_{\rm sb}, \delta_{\rm br}) = e^{-r^{\rm d}_{\rm A}/r_{0}} \triangleq \rho_{\rm c},
\end{align}
where $r_{0}$ is the decorrelation distance. Considering that the distance between sender and receiver ranges from $0$ to $2r^{\rm c}_{\rm A}$, we have $\rho_{\rm c} \in [0,e^{-2r^{\rm c}_{\rm A}/r_{0}}]$.

Since shadowing follows lognormal distribution with zero mean and $\sigma$~dB standard deviation, ${\delta}_{\rm sb}$ and ${\delta}_{\rm br}$ are correlated Gaussian variables that obey the following joint distribution,
\begin{align}
{f_{\rm p}}\left( {{{ \delta }_{{\rm{sb}}}},{{ \delta }_{{\rm{br}}}}} \right) = \frac{1}{{2\pi {\sigma ^2}\sqrt {1 - \rho _{\rm c}^2} }}\exp \left\{ { - \frac{{ \delta _{{\rm{sb}}}^2 - 2\rho_{\rm c} {{ \delta }_{{\rm{sb}}}}{{ \delta }_{{\rm{br}}}} +  \delta _{{\rm{br}}}^2}}{{2\left( {1 - \rho _{\rm c}^2} \right){\sigma ^2}}}} \right\}\label{eq:joint}.
\end{align}

From \eqref{eq:sna}, the network availability of the DF cellular mode can be expressed as follows,
\begin{align}
&\Pr\{\Pr \{ \mathcal{L}^{\rm c,II}|{\mu_{\rm sb}, \mu_{\rm br}}\}\leq {\varepsilon}_{\max}\}\nonumber\\
&= {\mathbb{E}}_{\delta_{\rm sb}, \delta_{\rm br}}\{{\mathbbm{1}}\left(\Pr\{\mathcal{L}^{\rm c,II}|r^{\rm c}_{\rm A}, \delta_{\rm sb}, \delta_{\rm br}\} \leq {\varepsilon}_{\max} \right)\}\nonumber\\
& = \int_{ - \infty }^{ + \infty } {\int_{ - \infty }^{ + \infty } {{{{\mathbbm{1}}}}\left( {\Pr \left\{ {{\mathcal{L}^{{\rm{c,II}}}}\big{|}r_{\rm{A}}^{\rm{c}},{{ \delta }_{{\rm{sb}}}},{{ \delta }_{{\rm{br}}}}} \right\} \le {\varepsilon _{\max }}} \right){f_{\rm{p}}}\left( {{{ \delta }_{{\rm{sb}}}},{{ \delta }_{{\rm{br}}}}} \right)} } {\rm d}{{ \delta }_{{\rm{sb}}}}{\rm d}{{ \delta }_{{\rm{br}}}}.\label{eq:int}
\end{align}
$\Pr \left\{ {{\mathcal{L}^{{\rm{c,II}}}}{\big{|}}r_{\rm{A}}^{\rm{c}},{{ \delta }_{{\rm{sb}}}},{{ \delta }_{{\rm{br}}}}} \right\} = {\varepsilon^{\rm c, II} _{\rm{sb}}} + {\varepsilon^{\rm c, II} _{\rm{br}}} - {\varepsilon^{\rm c, II} _{\rm{sb}}}  {\varepsilon^{\rm c, II} _{\rm{br}}}$ strictly decreases with  ${ \delta }_{{\rm{sb}}}$ and ${ \delta }_{{\rm{br}}}$. For any given  ${ \delta }_{{\rm{br}}}$ that satisfies ${\varepsilon^{\rm c, II} _{\rm{br}}} < {\varepsilon _{\max }}$, the value of  ${ \delta }_{{\rm{sb}}}$ that satisfies ${\Pr \left\{ {{\mathcal{L}^{{\rm{c,II}}}}\big{|}r_{\rm{A}}^{\rm{c}},{{ \delta }_{{\rm{sb}}}},{{ \delta }_{{\rm{br}}}}} \right\} = {\varepsilon _{\max }}}$ is unique, and we denote it as $F^{\rm c}_{\delta}(r_{\rm{A}}^{\rm{c}},{ \delta }_{{\rm{br}}})$. Then, ${\Pr \left\{ {{\mathcal{L}^{{\rm{c,II}}}}\big{|}r_{\rm{A}}^{\rm{c}},{{ \delta }_{{\rm{sb}}}},{{ \delta }_{{\rm{br}}}}} \right\} < {\varepsilon _{\max }}}$ if and only if ${ \delta }_{{\rm{sb}}} > F^{\rm c}_{\delta}(r_{\rm{A}}^{\rm{c}}, { \delta }_{{\rm{br}}})$. Therefore, \eqref{eq:int} can be re-expressed as follows,
\begin{align}
\int_{ { \delta }_{{\rm{br}}}^{\rm lb}}^{ + \infty } {\int_{ F^{\rm c}_{\delta}(r_{\rm{A}}^{\rm{c}},{ \delta }_{{\rm{br}}}) }^{ + \infty } {f_{\rm{p}}}\left( {{{ \delta }_{{\rm{sb}}}},{{ \delta }_{{\rm{br}}}}} \right)} {\rm d}{{ \delta }_{{\rm{sb}}}}{\rm d}{{ \delta }_{{\rm{br}}}},\label{eq:sint}
\end{align}
where ${ \delta }_{{\rm{br}}}^{\rm lb}$ is the value of ${ \delta }_{{\rm{br}}}$ that satisfies ${\varepsilon^{\rm c, II} _{\rm{br}}} = {\varepsilon _{\max }}$. If ${ \delta }_{{\rm{br}}} < { \delta }_{{\rm{br}}}^{\rm lb}$, then ${\varepsilon^{\rm c, II} _{\rm{br}}} > {\varepsilon _{\max }}$ and $\Pr \left\{ {{\mathcal{L}^{{\rm{c,II}}}}{\big{|}}r_{\rm{A}}^{\rm{c}},{{ \delta }_{{\rm{sb}}}},{{ \delta }_{{\rm{br}}}}} \right\} \leq {\varepsilon _{\max }}$ cannot be satisfied.

Similar to D2D mode, the maximal available range of the DF cellular mode can be obtained by solving the following problem,
\begin{align}
\mathop {\max }\limits_{{T_1},{T_2}} & \quad r_{\rm A}^{\rm c}\label{eq:aimCII}\\
\text{s.t.}\quad & 10\log_{10}{\mu _{{\rm{sb}}}} = -10\alpha\log_{10}{(r_{\rm{A}}^{\rm{c}})} + \delta _{\rm{sb}} + {\mu _0}, \label{eq:muCsb}\tag{\theequation a}\\
& 10\log_{10}{\mu _{{\rm{br}}}} = -10\alpha\log_{10}{(r_{\rm{A}}^{\rm{c}})} + \delta _{\rm{br}} + {\mu _0}, \label{eq:muCbr}\tag{\theequation b}\\
&\int_{ { \delta }_{{\rm{br}}}^{\rm lb} }^{ + \infty } {\int_{ F^{\rm c}_{\delta}(r_{\rm{A}}^{\rm{c}},{ \delta }_{{\rm{br}}}) }^{ + \infty } {f_{\rm{p}}}\left( {{{ \delta }_{{\rm{sb}}}},{{ \delta }_{{\rm{br}}}}} \right)} {\rm d}{{ \delta }_{{\rm{sb}}}}{\rm d}{{ \delta }_{{\rm{br}}}} \geq P_{\rm A}, \label{eq:PACII}\tag{\theequation c}\\
&T_1+T_2 \leq D_{\rm t}, \label{eq:Dt}\tag{\theequation d}\\
&T_1=k_1T_{\rm f}, T_2=k_2T_{\rm f}, k_1,k_2\in {\mathbb{Z}},\nonumber
\end{align}
where constraint \eqref{eq:PACII} can ensure the availability requirement.

\subsection{DF Multi-connectivity Mode} For the DF multi-connectivity mode, the problem is more complex than problem \eqref{eq:aimCII}. To maximize available range of this mode, we need some approximations on shadowing. The available range of D2D link cannot be too large. With small $r^{\rm d}_{\rm A}$, $\rho(\delta_{\rm sb}, \delta_{\rm br}) \approx 1$, and hence $\delta_{\rm sb}$ and $\delta_{\rm br}$ are approximately identical.\footnote{We will show that the correlation of shadowing between UL channel and DL channel has little impact on available range of cellular links via numerical results. Intuitively, with high correlation, the two transmissions are likely to fail at the same time. However, failures in both UL and DL do not make the situation worse than only one failure in UL or DL, because the packet is lost when either UL or DL fails.} When $r_{\rm sb}=r^{\rm c}_{\rm A}=r_{\rm br}$ and $\delta_{\rm sb} \approx \delta_{\rm br} \triangleq \delta_{\rm c}$, we have $\mu_{\rm sb} \approx \mu_{\rm br} \triangleq \mu_{\rm c}$. Since $r_{\rm sb}=r^{\rm c}_{\rm A}=r_{\rm br}$ and $\delta_{\rm sb} \approx \delta_{\rm br}$, the correlation coefficients of shadowing between cellular links and D2D link can be simplified as $\rho(\delta_{\rm sb}, \delta_{\rm sr}) = \rho(\delta_{\rm br}, \delta_{\rm sr} ) \triangleq \rho_{\rm d}$. From \eqref{eq:sna}, the network availability can be expressed as follows,
\begin{align}
&\Pr\{\Pr \{ {\mathcal{L}^{\rm{m,II}}}|{\mu_{\rm c}, \mu_{\rm sr}}\}\leq {\varepsilon}_{\max}\} \nonumber\\
& = \int_{ - \infty }^{ + \infty } {\int_{ - \infty }^{ + \infty } {{{{\mathbbm{1}}}}\left( {\Pr \left\{ {{\mathcal{L}^{{\rm{m,II}}}}\big{|}r^{\rm c}_{\rm A}, r_{\rm{A}}^{\rm{d}},{{ \delta }_{{\rm{c}}}},{{ \delta }_{{\rm{sr}}}}} \right\} \le {\varepsilon _{\max }}} \right){f_{\rm{p}}}( {{{ \delta }_{{\rm{c}}}}},{{ \delta }_{{\rm{sr}}}} )} } {\rm d}{{ \delta }_{{\rm{c}}}}{\rm d}{{ \delta }_{{\rm{sr}}}},\label{eq:Hint}
\end{align}
where ${f_{\rm{p}}}( {{{ \delta }_{{\rm{c}}}}},{{ \delta }_{{\rm{sr}}}} )$ is the joint PDF of ${\delta}_{\rm sr}$ and ${{ \delta }_{{\rm{sr}}}}$. Network availability in \eqref{eq:Hint} depends on both $r^{\rm c}_{\rm A}$ and $r_{\rm{A}}^{\rm{d}}$. In what follows, we fixed the available range of cellular links as the radius of each cell $r^{\rm c}_{\rm A} = R_{\rm cell}$, and maximize the available range of D2D links. The impacts of radius of cells on the available range of D2D links will be studied in Section VI.

For any given  ${ \delta }_{{\rm{sr}}}$, the value of  ${ \delta }_{{\rm{c}}}$ that satisfies $\Pr \left\{ {{\mathcal{L}^{{\rm{m,II}}}}\big{|}R_{\rm cell}, r_{\rm{A}}^{\rm{d}},{{ \delta }_{{\rm{c}}}},{{ \delta }_{{\rm{sr}}}}} \right\} = {\varepsilon _{\max }}$ is denoted as $F^{\rm m}_{\delta}(r_{\rm{A}}^{\rm{d}},{ \delta }_{{\rm{sr}}})$.
Then, \eqref{eq:Hint} can be simplified as follows,	
\begin{align}
\int_{ - \infty }^{ + \infty } {\int_{ F^{\rm m}_{\delta}(r_{\rm{A}}^{\rm{d}},{ \delta }_{{\rm{sr}}}) }^{ + \infty } {f_{\rm{p}}}( {{{ \delta }_{{\rm{c}}}}},{{ \delta }_{{\rm{sr}}}} ){\rm d}{{ \delta }_{{\rm{c}}}} } {\rm d}{{ \delta }_{{\rm{sr}}}}.\label{eq:sHint}
\end{align}

The maximal available range of D2D link can be obtained by solving the following problem,
\begin{align}
\mathop {\max }\limits_{{T_1},{T_2}} & \quad r_{\rm A}^{\rm d}\label{eq:aimHII}\\
\text{s.t.}\quad & 10\log_{10}{\mu _{{\rm{c}}}} = -10\alpha\log_{10}{(R_{\rm cell})} + \delta _{\rm{c}} + {\mu _0}, \label{eq:muHsb}\tag{\theequation a}\\
& 10\log_{10}{\mu _{{\rm{sr}}}} = - 10\alpha\log_{10}{(r_{\rm{A}}^{\rm{d}}) } + \delta _{\rm{sr}} + {\mu _0}, \label{eq:muHbr}\tag{\theequation b}\\
&\int_{ - \infty }^{ + \infty } {\int_{ F^{\rm m}_{\delta}(r_{\rm{A}}^{\rm{d}},{ \delta }_{{\rm{sr}}}) }^{ + \infty } {f_{\rm{p}}}( {{{ \delta }_{{\rm{c}}}}},{{ \delta }_{{\rm{sr}}}} ){\rm d}{{ \delta }_{{\rm{c}}}}} {\rm d}{{ \delta }_{{\rm{sr}}}}\geq P_{\rm A}, \label{eq:PAHII}\tag{\theequation c}\\
&\eqref{eq:Dt}, T_1=k_1T_{\rm f}, T_2=k_2T_{\rm f}, k_1,k_2\in {\mathbb{Z}}.\nonumber
\end{align}

\subsection{Algorithms for Solving Problems \eqref{eq:aimCII} and \eqref{eq:aimHII}}
The number of solutions of $T_1$ and $T_2$ that satisfy $\eqref{eq:Dt}$ and $T_1=k_1T_{\rm f}, T_2=k_2T_{\rm f}, k_1,k_2\in {\mathbb{Z}}$ is small. Hence, the problems can be solved by exhaustive search method. However, constraints \eqref{eq:PACII} and \eqref{eq:PAHII} are not in closed-form, we need compute them numerically. Since \eqref{eq:PACII} and \eqref{eq:PAHII} are similar, we take \eqref{eq:PAHII} as an example to illustrate how to compute the integration.

For any given $T_1$ and $T_2$, we need to find the maximal value of $r_{\rm A}^{\rm d}$ that satisfies \eqref{eq:PAHII}. For this purpose, we first prove the following property.
\begin{pro}\label{P:propertyRA}
\emph{The left hand side of \eqref{eq:PAHII} strictly decreases with $r_{\rm A}^{\rm d}$.}
\begin{proof}
See proof in Appendix \ref{App:Ploss}.
\end{proof}
\end{pro}
Property \eqref{P:propertyRA} indicates that the maximal value of $r_{\rm A}^{\rm d}$ is unique and is obtained when equality in \eqref{eq:PAHII} holds. Therefore, the maximal value of $r_{\rm A}^{\rm d}$ can be obtained via binary searching \cite{boyd}.

For a given value of $r_{\rm A}^{\rm d}$, \eqref{eq:PAHII} can be obtained with the following method. By replacing ${{ \delta }_{{\rm{sb}}}}$ and ${{ \delta }_{{\rm{br}}}}$ with ${{{ \delta }_{{\rm{c}}}}}$ and ${{ \delta }_{{\rm{sr}}}}$, we can obtain ${f_{\rm{p}}}( {{{ \delta }_{{\rm{c}}}}},{{ \delta }_{{\rm{sr}}}} )$ from  ${f_{\rm p}}\left( {{{ \delta }_{{\rm{sb}}}},{{ \delta }_{{\rm{br}}}}} \right)$ in  \eqref{eq:joint}. Then, by substituting ${f_{\rm{p}}}( {{{ \delta }_{{\rm{c}}}}},{{ \delta }_{{\rm{sr}}}} )$ into the left hand side of \eqref{eq:PAHII}, we can derive that
\begin{align}
&\int_{ -\infty }^{ + \infty } {\int_{ F^{\rm m}_{\delta}(r_{\rm{A}}^{\rm{d}},{ \delta }_{{\rm{sr}}}) }^{ + \infty } \frac{1}{{2\pi {\sigma ^2}\sqrt {1 - \rho _{\rm d}^2} }}\exp \left\{ { - \frac{{ \delta _{{\rm{c}}}^2 - 2\rho_{\rm d} {{ \delta }_{{\rm{c}}}}{{ \delta }_{{\rm{sr}}}} +  \delta _{{\rm{sr}}}^2}}{{2\left( {1 - \rho _{\rm d}^2} \right){\sigma ^2}}}} \right\}} {\rm d}{{ \delta }_{{\rm{c}}}}{\rm d}{{ \delta }_{{\rm{sr}}}} \nonumber\\
=&\int_{ -\infty}^{ + \infty } {\frac{1}{{\sqrt {2\pi } \sigma }}\exp \left\{ { - \frac{{ \delta _{{\rm{sr}}}^2}}{{2{\sigma ^2}}}} \right\}{f_{\rm{Q}}}\left( {\frac{{F_\delta ^{\rm{m}}(r_{\rm{A}}^{\rm{d}},{{ \delta }_{{\rm{sr}}}}) - {\rho _{\rm{d}}}{{ \delta }_{{\rm{sr}}}}}}{{\sigma \sqrt {1 - \rho _{\rm{d}}^2} }}} \right)} {\rm{d}}{{ \delta }_{{\rm{sr}}}}\label{eq:intHII},
\end{align}
where $f_{\rm Q}(\cdot)$ is the Q-function.

It is not hard to see that the packet loss probability $\Pr \{ {{\mathcal{L}^{{\rm{m,II}}}}{|}R_{\rm cell}, r_{\rm{A}}^{\rm{d}},{{ \delta }_{{\rm{c}}}},{{ \delta }_{{\rm{sr}}}}} \}$ in \eqref{eq:loss} strictly decreases with shadowing ${ \delta }_{\rm c}$ and ${ \delta }_{\rm sr}$. For any given value of ${ \delta }_{{\rm{sr}}}$, the value of $F^{\rm m}_{\delta}(r_{\rm{A}}^{\rm{d}},{ \delta }_{{\rm{sr}}})$ (i.e., ${{ \delta }_{{\rm{c}}}}$ that satisfies $\Pr \{ {{\mathcal{L}^{{\rm{m,II}}}}{|}R_{\rm cell}, r_{\rm{A}}^{\rm{d}},{{ \delta }_{{\rm{c}}}},{{ \delta }_{{\rm{sr}}}}} \} = {\varepsilon _{\max }}$) can be obtained via binary searching. The algorithm for maximizing available range of the DF multi-connectivity mode is shown in Table~I.

Our algorithm uses binary searching for a given $T_1$. Denote the complexity of the binary searching as $\kappa$, which is very low \cite{boyd}. The complexity of the algorithm depends on the searching space of $T_1$, and can be expressed as $O(\frac{D_{\max}}{T_1}\kappa)$. When $D_{\max}$ is short, the complexity is low. It is ture that the complexity for computing $F^{\rm m}_{\delta}(r_{\rm{A}}^{\rm{d}},{ \delta }_{{\rm{sr}}})$ is high. However, we can compute it offline, and hence it will not lead to extra delay for URLLC.

\renewcommand{\algorithmicrequire}{\textbf{Input:}}
\renewcommand{\algorithmicensure}{\textbf{Output:}}
\begin{table}[htb]\small
	\caption{Algorithm for Maximizing Available Range}
	\vspace{-0.6cm}
	\begin{tabular}{p{16cm}}
		\\\hline
	\end{tabular}
	\begin{algorithmic}[1]
		\REQUIRE Number of users $K$, total bandwidth $W_{\max}$, number of antennas at the BS $N_{\rm t}$, packet size $b$, noise spectral density $N_0$, frame duration $T_{\rm f}$, E2E delay $D_{\max}$, and overall packet loss requirement ${\varepsilon _{\max }}$.
		\ENSURE Transmission durations $T_1$, $T_2$ and available range $r_{\rm A}^{\rm d}$.
		\STATE Set $l := 1$ and $T(l) := T_{\rm f}$.
		\WHILE{$T(l) < D_{\max}$}
		\STATE Set $T_1 := T(l)$ and $T_2 := D_{\max} - T(l)$.
		\STATE Set an initial value of communication range $r_0$.
		\IF {$\int_{ - \infty }^{ + \infty }f_{\rm Q}\left(F^{\rm m}_{\delta}(r_0,{ \delta }_{{\rm{sr}}})/\sigma\right){f_{\rm{p}}}({{ \delta }_{{\rm{sr}}}}){\rm d}{{ \delta }_{{\rm{sr}}}} < P_{\rm A}$}
		\STATE Set $r_{\rm lb} := r_0$
		\WHILE{$\int_{ - \infty }^{ + \infty }f_{\rm Q}\left(F^{\rm m}_{\delta}(r_0,{ \delta }_{{\rm{sr}}})/\sigma\right){f_{\rm{p}}}({{ \delta }_{{\rm{sr}}}}){\rm d}{{ \delta }_{{\rm{sr}}}} < P_{\rm A}$}
		\STATE Set $r_0 := 2r_0$.
		\ENDWHILE
		\STATE Set $r_{\rm ub} := r_0$.
		\ELSE
		\STATE Set $r_{\rm ub} := r_0$.
		\WHILE{$\int_{ - \infty }^{ + \infty }f_{\rm Q}\left(F^{\rm m}_{\delta}(r_0,{ \delta }_{{\rm{sr}}})/\sigma\right){f_{\rm{p}}}({{ \delta }_{{\rm{sr}}}}){\rm d}{{ \delta }_{{\rm{sr}}}} \geq P_{\rm A}$}
		\STATE Set $r_0 := r_0/2$.
		\ENDWHILE
		\STATE Set $r_{\rm lb} := r_0$.
		\ENDIF
		\STATE Find $r(l) \in [r_{\rm lb},r_{\rm ub}]$ that satisfies $\int_{ - \infty }^{ + \infty }f_{\rm Q}\left(F^{\rm m}_{\delta}(r(l),{ \delta }_{{\rm{sr}}})/\sigma\right){f_{\rm{p}}}({{ \delta }_{{\rm{sr}}}}){\rm d}{{ \delta }_{{\rm{sr}}}} = P_{\rm A}$ via binary searching.
		\STATE Set $T(l+1) := T(l) + T_{\rm f}$ and $l := l + 1$.
		\ENDWHILE
		\STATE $l^* := \arg \mathop {\max }\limits_{l} r{(l)}$.
		\RETURN $T_1 = T(l^*)$, $T_2 = D_{\max} - T_1$ and $r_{\rm A}^{\rm d} = r{(l^*)}$.
	\end{algorithmic}
	\vspace{-0.4cm}
	\begin{tabular}{p{16cm}}
		\\
		\hline
	\end{tabular}
	\vspace{-0.2cm}
\end{table}

%
%
%

\section{Comparison of Available Ranges}
In this section, we ignore processing delay and backhaul delay (i.e., $D_{\rm p}=D_{\rm_b}=0$ and $D_{\rm t} = D_{\max}$) and compare the available ranges of different transmission modes. The impact of processing delay on the available ranges of multi-connectivity modes will be shown in the next section. For fair comparison, we set $T_1 = T_2 = D_{\rm t}/2$ for all the modes.

Without closed-form expressions of available ranges, it is very challenging to analyze available ranges directly. Alternatively, we fix $\mu_{\rm{sb}}$, $\mu_{\rm{br}}$ and $\mu _{\rm{sr}}$ (i.e., fixing the communication distances and shadowing), and compare the packet loss probabilities of different modes. This is because the transmission modes with smaller packet loss probabilities can achieve larger available ranges under the same constraint on packet loss probability.

\subsection{AF Multi-connectivity Mode versus D2D Mode and AF Cellular Mode}
Since $\gamma^{\rm m, I}$ in \eqref{eq:SNRHI} is always higher than $\gamma^{\rm c,I}$ in \eqref{eq:SNRC} for any given $\mu_{\rm{sb}}$, $\mu_{\rm{br}}$ and $\mu _{\rm{sr}}$, the packet loss probability with the AF multi-connectivity mode is smaller than the packet loss probability with the AF cellular mode. In other words, for given communication distances, if the reliability requirement can be satisfied with the AF cellular mode, it can also be satisfied with the AF multi-connectivity mode. Therefore, the available range of the AF cellular mode does not exceed the available range of the AF multi-connectivity mode.


To compare the packet loss probabilities with the D2D mode and the AF multi-connectivity mode, we first study the relation between $\gamma^{\rm m, I}$ and $\mu_{\rm br}$, which is shown by the following property,

\begin{pro}\label{P:propertySNR2}
	\emph{$\gamma^{\rm m, I}$ lies in between $\beta $ and $ \frac{{{\mu _{{\rm{sr}}}}{g_{{\rm{sr}}}}P_{\rm{s}}^{\rm{t}}}}{{{N_0}W}}$. If $\beta < \frac{{{\mu _{{\rm{sr}}}}{g_{{\rm{sr}}}}P_{\rm{s}}^{\rm{t}}}}{{{N_0}W}}$, $\gamma^{\rm m, I}$ decreases with $\mu_{\rm br}$. Otherwise, $\gamma^{\rm m, I}$ increases with $\mu_{\rm br}$.}
	\begin{proof}
		See proof in Appendix \ref{App:mu2}.
	\end{proof}
\end{pro}
For the case $\beta <  \frac{{{\mu _{{\rm{sr}}}}{g_{{\rm{sr}}}}P_{\rm{s}}^{\rm{t}}}}{{{N_0}W}}$, forwarding received signal and noise leads to low receive SNR. Then, the packet loss probability with the AF multi-connectivity mode is higher than the packet loss probability with the D2D mode. Otherwise, the AF multi-connectivity mode outperforms the D2D mode.


\subsection{DF Multi-connectivity Mode versus D2D Mode and DF Cellular Mode}
In the second phase, $\gamma^{\rm m,II}$ in \eqref{eq:SNRH2} is the sum of the SNR with the D2D and the DF cellular modes, and is higher than the SNR only with the D2D mode or the DF cellular mode. Therefore, ${\varepsilon_2^{\rm{m,II}}}$ in \eqref{eq:eDL} is smaller than ${\varepsilon^{\rm d}_{\rm{sr}}}$ and ${\varepsilon^{\rm c,II}_{\rm{br},2}}$. From \eqref{eq:loss} and ${\varepsilon_2^{\rm{m,II}}} < {\varepsilon^{\rm d}_{\rm{sr}}}$, we can derive that
\begin{align}
\Pr\{\mathcal{L}^{\rm m,II}|\mu_{\rm{sb}},\mu_{\rm{br}},\mu _{\rm{sr}}\} & = \varepsilon^{\rm d}_{{\rm sr,}1}(1 - {\varepsilon^{\rm c, II} _{\rm{sb}}}){\varepsilon_2^{\rm{m,II}}} + \varepsilon^{\rm d}_{{\rm sr,}1} {\varepsilon^{\rm c, II} _{\rm{sb}}}\varepsilon^{\rm d}_{{\rm sr,}2}\nonumber\\
& <  \varepsilon^{\rm d}_{{\rm sr,}1}(1 - {\varepsilon^{\rm c, II} _{\rm{sb}}}){{\varepsilon^{\rm d}_{\rm{sr},2}}} + \varepsilon^{\rm d}_{{\rm sr,}1} {\varepsilon^{\rm c, II} _{\rm{sb}}}\varepsilon^{\rm d}_{{\rm sr,}2}\nonumber\\
&=\varepsilon^{\rm d}_{{\rm sr,}1}{{\varepsilon^{\rm d}_{\rm{sr},2}}},\label{eq:HD}
\end{align}
which is the same as $\Pr\{\mathcal{L}^{\rm d}|\mu _{\rm{sr}}\}$ in \eqref{eq:D2D}.

Moreover, from \eqref{eq:loss} and ${\varepsilon_2^{\rm{m,II}}} < {\varepsilon^{\rm c,II}_{\rm{br},2}}$, we can derive that
\begin{align}
\Pr\{\mathcal{L}^{\rm m,II}|\mu_{\rm{sb}},\mu_{\rm{br}},\mu _{\rm{sr}}\} & = \varepsilon^{\rm d}_{{\rm sr,}1}(1 - {\varepsilon^{\rm c, II} _{\rm{sb}}}){\varepsilon_2^{\rm{m,II}}} + \varepsilon^{\rm d}_{{\rm sr,}1} {\varepsilon^{\rm c, II} _{\rm{sb}}}\varepsilon^{\rm d}_{{\rm sr,}2}\nonumber\\
& <  \varepsilon _{{\rm{sr}},1}^{\rm{d}}(1 - \varepsilon _{{\rm{sb}}}^{{\rm{c}},{\rm{II}}})\varepsilon _{{\rm{br}}}^{{\rm{c}},{\rm{II}}} + \varepsilon _{{\rm{sr}},1}^{\rm{d}}\varepsilon _{{\rm{sb}}}^{{\rm{c}},{\rm{II}}}\varepsilon _{{\rm{sr}},2}^{\rm{d}}\nonumber\\
&\leq \varepsilon _{{\rm{br}}}^{{\rm{c}},{\rm{II}}} - \varepsilon _{{\rm{sb}}}^{{\rm{c}},{\rm{II}}}\varepsilon _{{\rm{br}}}^{{\rm{c}},{\rm{II}}}+ \varepsilon _{{\rm{sb}}}^{{\rm{c}},{\rm{II}}},\label{eq:HC}
\end{align}
which is the same as $\Pr\{\mathcal{L}^{\rm c,II}|\mu_{\rm sb},\mu_{\rm br}\}$ in \eqref{eq:CII}.

The results in \eqref{eq:HD} and \eqref{eq:HC} indicate that the available range of the DF multi-connectivity mode is larger than the available range with the D2D and DF cellular modes.

\subsection{Comparison of the Multi-connectivity Modes}
If the received SNR in the UL transmission is lower than the D2D transmission, according to the previous analysis, available range achieved by the D2D mode is larger than the AF multi-connectivity mode and smaller than the DF multi-connectivity mode. Therefore, the DF multi-connectivity mode can achieve larger available range than the AF multi-connectivity mode.

If the received SNR in the UL transmission is higher than the D2D transmission, then the BS can decode the packet with high probability (higher than the D2D mode). From \eqref{eq:SNRHI} and \eqref{eq:SNRH2}, $\gamma^{\rm m,I}$ achieved by the AF hyrbid mode is smaller than $\gamma^{\rm m,II}$ achieved by the DF hyrbid mode, and hence the DF multi-connectivity mode can achieve larger available range than the AF multi-connectivity mode.

If the received SNR in the UL transmission goes to infinity, we can derive that
\begin{align}
\mathop {\lim }\limits_{\beta \to \infty } \gamma^{\rm m,I}=\frac{g_{\rm br}}{N_{\rm t}}\frac{{{\mu _{{\rm{br}}}}P_{\rm{b}}^{\rm{t}} + {\mu _{{\rm{sr}}}}{g_{{\rm{sr}}}}P_{\rm{s}}^{\rm{t}}}}{{{N_0}W}}\label{eq:gammalim}.
\end{align}
With multiple antennas, $\frac{g_{\rm br}}{N_{\rm t}} \approx 1$. By substituting the approximation into \eqref{eq:SNRH2} and \eqref{eq:gammalim}, we can obtain that \eqref{eq:SNRH2} and \eqref{eq:gammalim} are the same. Thus, the available range of the AF multi-connectivity mode can approach the available range with the DF multi-connectivity mode when the UL SNR is high.

\section{Simulation and Numerical Results}
In this section, we first validate the approximation in \eqref{eq:SIMO} via simulation. Then, we illustrate the optimized available ranges of the cellular and multi-connectivity modes, respectively. Moreover, we show the performance gain of the multi-connectivity modes compared with the D2D and cellular modes. Finally, the impact of the processing delay on the available ranges of the multi-connectivity modes is demonstrated. The path loss model is $35.3+37.6 \log_{10}\{d~(\text{m})\}$. Other parameters are listed in Table II, unless otherwise specified.

\begin{table}[htbp]
\vspace{-0.4cm}\small
\renewcommand{\arraystretch}{1.3}
\caption{Parameters}
\begin{center}\vspace{-0.2cm}
\begin{tabular}{|p{5cm}|p{2.5cm}||p{4.5cm}|p{2cm}|}
  \hline
  E2D Delay $D_{\max}$ & $1$~ms \cite{3GPP2016Scenarios}&
  Duration of each frame $T_{\rm f}$ & $0.1$~ms \cite{3GPP2017NR} \\\hline
  One hop backhaul latency $D_{\rm b}$& $0.1$~ms \cite{Gongzheng2016Backhaul}&
  Processing delay at BS $D_{\rm p}$ & $0.1 \sim 0.7$~ms\\\hline
  Network availability $P_{\rm A}$ & $99.999$~\% \cite{Popovski2014METIS}&
  Packet loss probability $\varepsilon _{\max}$ & $10^{-7}$ \cite{Gerhard2014The} \\\hline
  Inter-site distance & $500$~m  \cite{3GPP2010Shadowing} &
  Total bandwidth & $20$~MHz  \cite{3GPP2010Shadowing} \\\hline
  Total transmit power of the BS & $46$~dBm \cite{3GPP2010Shadowing}  &
  Transmit power of each user & $23$~dBm  \cite{3GPP2010Shadowing}\\\hline
  Single-sided noise spectral density $N_0$ & $-173$ dBm/Hz \cite{3GPP2010Shadowing} &
  Decorrelation distance of shadowing $r_0$ & $100$~m \cite{WirelessCom}  \\\hline
  Standard deviation of shadowing & $8$ dB \cite{3GPP2010Shadowing} &
  Packet size $b$ & $20$~bytes \cite{3GPP2016Scenarios}
  \\\hline

\end{tabular}
\end{center}
\vspace{-0.5cm}
\end{table}

\begin{figure}[htbp]
	\vspace{-0.3cm}
	\centering
	\begin{minipage}[t]{0.6\textwidth}
		\includegraphics[width=1\textwidth]{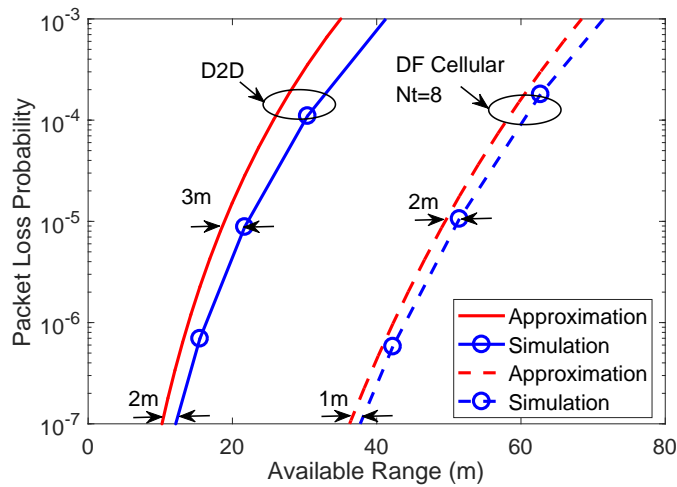}
	\end{minipage}
	\vspace{-0.2cm}
	\caption{Impacts of the approximation in \eqref{eq:SIMO} on the available range.}
	\label{fig:Approx}
	\vspace{-0.2cm}
\end{figure}

To validate the accuracy of the approximation in \eqref{eq:SIMO}, we show its impact on the available range in Fig. \ref{fig:Approx}. The lognormal distributed shadowing is considered. To satisfy the network availability requirement, the shadowing of each link is set as the threshold $\delta^{\rm th}$ such that $\Pr\{\delta \geq \delta^{\rm th}\} = P_{\rm A}$. Bandwidth allocated to each sender is $2$~MHz (i.e., $20$~MHz bandwidth is shared by $10$ senders), and the transmission duration of each phase is $4T_{\rm f}$. The approximation results of packet loss probabilities with D2D and DF cellular modes are obtained from  from \eqref{eq:D2D} and \eqref{eq:CII}, respectively, where \eqref{eq:SIMO} is used to compute the decoding error probabilities of each transmission. In simulation, we generate the $10^{10}$ instantaneous channel gains of each link, and compute the decoding error probabilities via the left hand side of \eqref{eq:DE}. The results in Fig. \ref{fig:Approx} indicate that the gap between the simulation and approximation ranges from $1$ to $3$~m. Besides, the gap shrinks as the packet loss probabilities decrease or the number of antennas increases. Therefore, the approximation is very accurate for URLLC with multiple antennas.

The simulation results in Fig. \ref{fig:Approx} also indicate that there is a tradeoff between packet loss probability and available range. For example, if the required packet loss probability is around $10^{-3}$ (i.e., reliability requirement of traditional video and audio services), and the E2E delay is $1$~ms, then the available ranges of D2D and DF cellular modes are around $35$~m and $70$~m, respectively. However, for URLLC that requires $10^{-7}$ packet loss probability, the available ranges of D2D and DF cellular modes are around $10$~m and $35$~m, respectively. This observation implies that achieving satisfactory network availability under the stringent QoS constraints of URLLC is very challenging.


\begin{figure}[htbp]
	\vspace{-0.3cm}
	\centering
	\begin{minipage}[t]{0.6\textwidth}
		\includegraphics[width=1\textwidth]{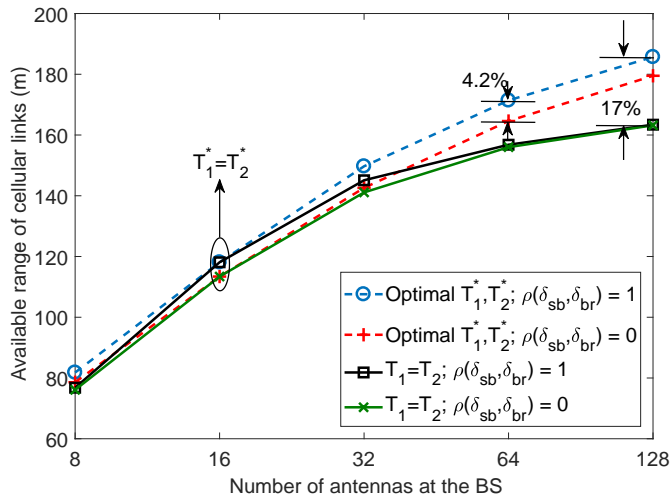}
	\end{minipage}
	\vspace{-0.2cm}
	\caption{Available range of cellular links (DF cellular mode) v.s. number of antennas at the BS, where $K=10$, $D_{\rm b} = 0$ and $D_{\rm p} = 0$.}
	\label{fig:CellModeII}
	\vspace{-0.2cm}
\end{figure}
The available ranges achieved by the DF cellular mode are illustrated in Fig. \ref{fig:CellModeII}, where the results with $T^*_1$ and $T^*_2$ are obtained by solving problem \eqref{eq:aimCII}. Considering that the DF cellular mode is better than the AF cellular mode without processing delay, by setting $D_{\rm b} = 0$ and $D_{\rm p} = 0$ in the DF cellular mode we obtain the upper bounds of the available range that can be achieved by the cellular modes. Even with $N_{\rm t} = 128$, the available range is less than the radius of macro cell (i.e., $250$m). Therefore, the available range is unsatisfactory  with cellular modes. Moreover, the results also indicate that correlation of shadowing in UL and DL has little impact on the available range (i.e., the gap between the available range when $\rho(\delta_{\rm sb},\delta_{\rm br}) = 1$ and the available range when $\rho(\delta_{\rm sb},\delta_{\rm br}) = 0$ is $4.6$\%.).
In the rest of this section, we set $\delta_{\rm sb}=\delta_{\rm br}=\delta_{\rm c}$ and study the correlation of $\delta_{\rm c}$ and $\delta_{\rm sr}$ on the available range.

\begin{figure}[htbp]
	\vspace{-0.3cm}
	\centering
	\begin{minipage}[t]{0.6\textwidth}
		\includegraphics[width=1\textwidth]{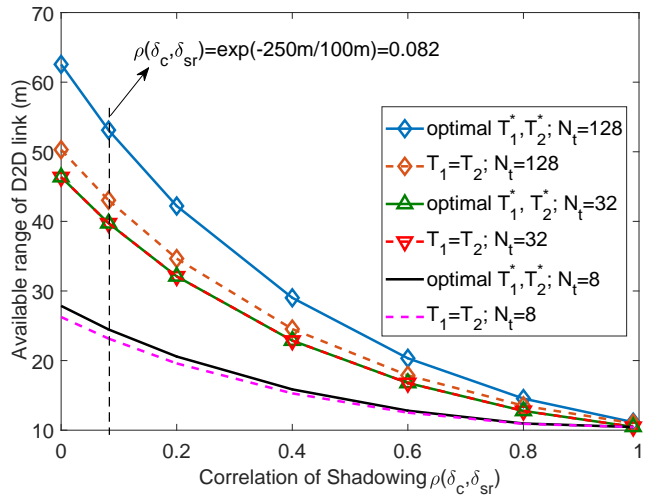}
	\end{minipage}
	\vspace{-0.2cm}
	\caption{Available range of D2D links (under DF multi-connectivity mode) v.s. correlation of shadowing between cellular and D2D links, where $D_{\rm p} = D_{\rm b} = T_{\rm f}$, $K = 10$ and $R_{\rm cell} = 250$~m.}
	\label{fig:correlation}
	\vspace{-0.2cm}
\end{figure}
The impact of $\rho(\delta_{\rm c}, \delta_{\rm sr})$ on $r^{\rm d}_{\rm A}$ under the DF multi-connectivity mode is shown in Fig. \ref{fig:correlation}. The optimal transmission durations $T^*_1$ and $T^*_2$ are obtained by solving problem \eqref{eq:aimHII}. Different from the results in Fig. \ref{fig:CellModeII}, $\rho(\delta_{\rm c}, \delta_{\rm sr})$ has significant impact on the available range of D2D links. This can be explained as follows. On the one hand, UL and DL transmissions of the DF cellular mode are arranged successively. The packet is lost if one of the transmissions fails. Even if both transmissions fail at the same time, the situation does not become worse since the packet is lost anyway. On the other hand, with the DF multi-connectivity mode, each packet is transmitted over a D2D link and a cellular link in parallel. The packet is lost when both the D2D and cellular links fail. As a result, the decoding error probability increases with the correlation of shadowing of D2D and cellular links rapidly. Therefore, if a packet is transmitted over links sequentially, then the available range is insensitive to the correlation of shadowing of the links. If a packet is transmitted over parallel links, then the available range decreases with the correlation of shadowing on different links significantly. For macro cell with a large radius $R_{\rm cell} = 250$~m, $\rho(\delta_{\rm c}, \delta_{\rm sr}) = 0.082$, the available range  decreases $20$\%. Finally, comparing to the case $T_1=T_2$, the performance gain by optimizing transmission durations in the two phases is presented in Fig. \ref{fig:correlation}. When $N_{\rm t} = 32$, $T^*_1 = T^*_2= 4T_{\rm f}$, which is the same as the configuration that simply set the durations of the two phases as equal. When $N_{\rm t} = 8$, $T^*_1 = 5T_{\rm f}$ and $T^*_2 = 3T_{\rm f}$, and the performance gain is trivial. This is because with small $N_{\rm t}$, the reliability of the cellular link is low compared with D2D links, and hence $T^*_1$ and $T^*_2$ are close to that in the D2D mode (i.e., $T_1=T_2$). When $N_{\rm t} = 128$, $T^*_1 = 2T_{\rm f}$ and $T^*_2 = 6T_{\rm f}$, and the performance gain is around $20$~\%. Since the received SNR in the UL transmission increases with $N_{\rm t}$ due to the array gain, the packet can be decoded successfully with a short transmission duration when $N_{\rm t}$ is large. By contrast, since BSs do not have CSI in DL transmission, there is no array gain. Therefore, more frames are allocated for the DL transmission.

\begin{figure}[htbp]
	\vspace{-0.3cm}
	\centering
	\begin{minipage}[t]{0.6\textwidth}
		\includegraphics[width=1\textwidth]{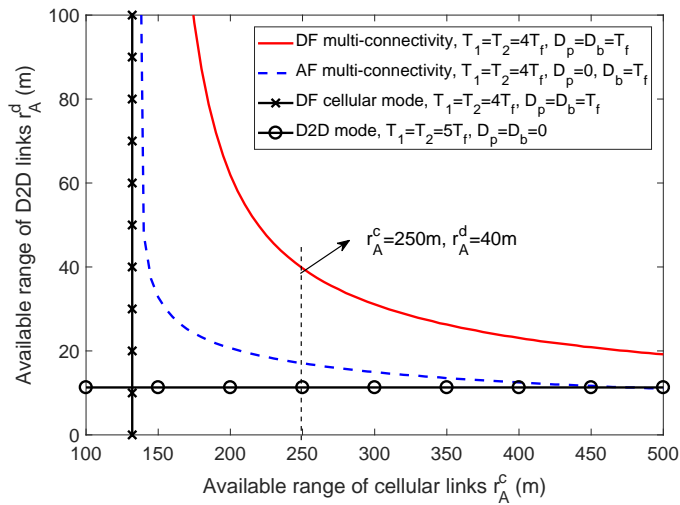}
	\end{minipage}
	\vspace{-0.2cm}
	\caption{Available range of D2D links v.s. available range of cellular links, where $K=10$ and $N_{\rm t} = 32$.}
	\label{fig:CellvsD2D}
	\vspace{-0.2cm}
\end{figure}

The relationship between the available range of D2D links and the available range of cellular links is illustrated in Fig. \ref{fig:CellvsD2D}. For a fair comparison, we set $T_1=T_2$ for all the modes. Similar to \eqref{eq:cor}, the correlation of $\delta_{\rm c}$ and $\delta_{\rm sr}$ is determined by $\rho(\delta_{\rm c}, \delta_{\rm br}) = e^{-r^{\rm c}_{\rm A}/r_{0}}$. If the communication distances of the D2D and cellular links are shorter than the available ranges of the D2D and cellular links, respectively, then the network availability and QoS requirement can be satisfied. For example, if we set the available range of the cellular links equal to the radius of the macro cell, i.e., $r^{\rm c}_{\rm A} = 250$~m, then for a pair of sender and receiver with distance shorter than $40$~m, the availability and QoS requirement can be satisfied with the DF multi-connectivity mode. The results indicate that there is a tradeoff between $r^{\rm d}_{\rm A}$ and $r^{\rm c}_{\rm A}$. With the DF multi-connectivity mode, the tradeoff is improved remarkably. Besides, the results indicate that the available range of D2D links increases as the radius of each BS decreases. This observation implies that network availability increases with the density of BSs.

\begin{figure}[htbp]
\vspace{-0.4cm}
\centering
\subfigure[{Available range of D2D links v.s. number of antennas, where $R_{\rm cell} = 250$~m and $\rho(\delta_{\rm c}, \delta_{\rm sr}) = \exp(-R_{\rm cell}/r_0) = 0.0821$.}]{
\label{fig:H1H2_Nt} 
\includegraphics[width=0.6\textwidth]{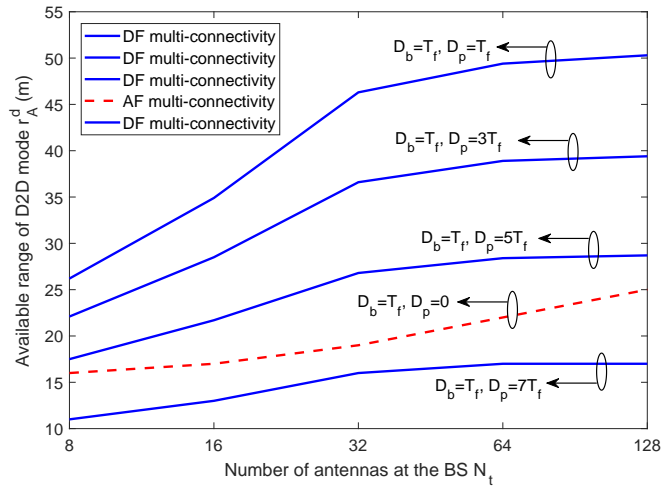}}
\vspace{-0.2cm}
\subfigure[{Available range of D2D links v.s. correlation of shadowing on cellular and D2D links, where $D_{\rm b} = T_{\rm f}$, $R_{\rm cell} = 100$~m and $N_{\rm t} = 8$.}]{
\label{fig:H1H2_Cor} 
\includegraphics[width=0.6\textwidth]{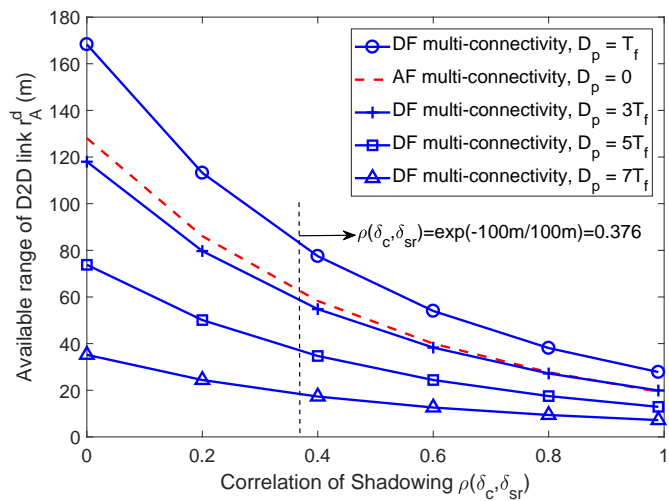}}
\caption{Available ranges with different processing delays.}
 \label{fig:ranges} 
\vspace{-0.6cm}
\end{figure}
The available ranges with different processing delays are shown in Fig. \ref{fig:ranges}. In Fig. \ref{fig:H1H2_Nt}, the available range of cellular links is set to be the radius of the macro cell, and the relation between available range and the number of antennas at the BS is provided. The curves are not smooth because the optimized durations of the two phases are discretized by frame duration. In Fig. \ref{fig:H1H2_Cor}, the available range of cellular links is set to be the radius of the micro cell (i.e., $R_{\rm cell} = 100$~m). When $r^{\rm c}_{\rm A}$ is short, the correlation of shadowing on cellular links and D2D links is high. To show the performance loss due to correlation of shadowing, the relation between available range and $\rho(\delta_{\rm c}, \delta_{\rm br})$ is provided.

As shown in Fig. \ref{fig:ranges}, when the processing delay is short (e.g., $D_{\rm p} = T_{\rm f}$), the DF multi-connectivity mode can achieve larger available range than the AF multi-connectivity mode. For macro cell, the radius is large, and hence the UL SNR is low. Since noise in the UL is amplified with the AF multi-connectivity mode, the DF multi-connectivity mode outperforms the AF multi-connectivity mode even when $D_{\rm p} \leq 5 T_{\rm f}$. However, the UL SNR is high for micro cell due to short radius, and hence the AF multi-connectivity mode outperforms the DF multi-connectivity mode when $D_{\rm p} \geq 3 T_{\rm f}$.


\section{Conclusion}
In this work, we established a framework for improving available ranges for URLLC with different transmission modes. The processing delay for decoding packets at the BSs, decoding errors in short blocklength regime and cross-correlation of shadowing were taken into consideration in our framework. The available ranges of DF modes were maximized by optimizing transmission durations of the two-phase protocol. Besides, we compared the available ranges of different transmission modes. Analytical results indicated that the DF multi-connectivity mode can achieve larger available ranges than other modes when the processing delay is small (e.g., $0.1 \sim 0.3$~ms). The AF multi-connectivity mode can approach the performance of the DF multi-connectivity mode when the receive SNR at the BS is high. Numerical results show that if two links are arranged in series, such as the UL and DL channels in cellular modes, the available range is insensitive to the correlation of shadowing. If two copies of one packet are transmitted in parallel links, the available range decreases rapidly as the shadowing correlation increases. The results also show that there is a tradeoff between available ranges of D2D and cellular links, and the tradeoff can be improved remarkably with the DF multi-connectivity mode. Besides, increasing the number of antennas at the BSs or the density of BSs are helpful for improving network availability.
\vspace{-0.1cm}

\appendices

\section{Decoding Error Probability in SIMO Systems \eqref{eq:SIMO}}
\label{App:SIMO}
\renewcommand{\theequation}{A.\arabic{equation}}
\setcounter{equation}{0}
\begin{proof}
To derive the decoding error probability, we first need to obtain the CDF of SNR. $F_{\gamma}(x)=\Pr\{\gamma \leq x\}=\Pr\{g \leq \frac{N_0Wx}{\mu_{\rm sr}P^{\rm t}_{\rm s}}\}$, where the PDF of $g$ is $f_g(x) = \frac{1}{(N_{\rm t}-1)!}x^{N_{\rm t}-1}\exp(-x)$. Denote $y = \frac{N_0Wx}{\mu_{\rm sr}P^{\rm t}_{\rm s}}$. Then, the CDF can be re-expressed as
$\Pr\{g \leq y\} = \int_0^y {\frac{{{x^{{N_{\rm{t}}}{\rm{ - }}1}}}}{{\left( {{N_{\rm{t}}} - 1} \right)!}}{e^{ - x}}{\rm d}x}  =  - \frac{{{x^{{N_{\rm{t}}}{\rm{ - }}1}}}}{{\left( {{N_{\rm{t}}} - 1} \right)!}}{e^{ - x}}\Big{|}_0^y + \int_0^y {\frac{{{x^{{N_{\rm{t}}}{\rm{ - 2}}}}}}{{\left( {{N_{\rm{t}}} - 2} \right)!}}{e^{ - x}}{\rm d}x}$. By setting $\tilde{A}_{n}(y) = \int_0^y {\frac{{{x^{n}}}}{{ {n} !}}{e^{ - x}}{\rm d}x}$, we can obtain that $\Pr\{g \leq y\} = \tilde{A}_{N_{\rm t}-1}(y) = - \frac{{{y^{{N_{\rm{t}}}{\rm{ - }}1}}}}{{\left( {{N_{\rm{t}}} - 1} \right)!}}{e^{ - y}} + \tilde{A}_{N_{\rm t}-2}(y)$.
Further considering that $\tilde{A}_0(y) = 1-\exp(-y)$, we have
\begin{align}
\int_0^y {\frac{{{x^{{N_{\rm{t}}}{\rm{ - }}1}}}}{{\left( {{N_{\rm{t}}} - 1} \right)!}}{e^{ - x}}{\rm d}x} = \tilde{A}_{N_{\rm t}-1}(y) = 1-\exp(-y)\sum_{n=0}^{N_{\rm t}-1}\frac{y^n}{n!}.\label{eq:AppCDF}
\end{align}
From \eqref{eq:AppCDF}, we can further derive that
\begin{align}
\int_\zeta ^\xi  {{F_\gamma }\left( x \right){\rm d}x} &= \int_{g_{\rm L}} ^{g_{\rm U}}  {\left[1-\exp(-y)\sum_{n=0}^{N_{\rm t}-1}\frac{y^n}{n!}\right] {\rm d}(\frac{\mu_{\rm sb}P^{\rm t}_{\rm s}}{N_0W}y)}\nonumber\\
&= \frac{\mu_{\rm sb}P^{\rm t}_{\rm s}}{N_0W}\left[(g_{\rm U}-g_{\rm L}) - \sum_{n=0}^{N_{\rm t}-1} \left(\int_{{0}}^{{g_{\rm{U}}}} {  \frac{{{y^n}}}{{n!}}e^{ - y}{\rm{d}}y}-\int_{{0}}^{{g_{\rm{L}}}} { \frac{{{y^n}}}{{n!}}e^{ - y}{\rm{d}}y} \right)\right],\label{eq:AppDecode}
\end{align}
where $g_{\rm U} = \frac{N_0W\xi}{\mu_{\rm sb}P^{\rm t}_{\rm s}}$ and $g_{\rm L} = \frac{N_0W\zeta}{\mu_{\rm sb}P^{\rm t}_{\rm s}}$. By applying the result in \eqref{eq:AppCDF}, \eqref{eq:AppDecode} can be expressed as
\begin{align}
\int_\zeta ^\xi  {{F_\gamma }\left( x \right){\rm d}x} = \frac{\mu_{\rm sb}P^{\rm t}_{\rm s}}{N_0W}\left[(g_{\rm U}-g_{\rm L}) - \sum_{n=0}^{N_{\rm t}-1} \left(N_{\rm t}-n \right)A_n\right],\label{eq:AppAn}
\end{align}
where $A_n = \frac{g_{\rm L}^n}{n!} e^{-g_{\rm L}} - \frac{g_{\rm U}^n}{n!}e^{-g_{\rm U}}$. By substituting \eqref{eq:AppAn} into \eqref{eq:DE}, we can obtain \eqref{eq:SIMO}.
\end{proof}

\section{CDF of SNR with DF Multi-Connectivity Mode \eqref{eq:CDFmII}}
\label{App:CDF}
\renewcommand{\theequation}{B.\arabic{equation}}
\setcounter{equation}{0}
\begin{proof}
For notational simplicity, we denote $C_{\rm br} = \frac{\mu_{\rm br}P^{\rm t}_{\rm b}}{N_0WN_{\rm t}}$ and $C_{\rm sr} = \frac{\mu_{\rm br}P^{\rm t}_{\rm s}}{N_0W}$. Then, \eqref{eq:SNRH2} can be simplified as $\gamma^{\rm m,II} = C_{\rm br}g_{\rm br}+C_{\rm sr}g_{\rm sr}$. The CDF of SNR can be expressed as $F_{{\gamma}^{\rm m,II}}(x) = \Pr\{C_{\rm br}g_{\rm br}+C_{\rm sr}g_{\rm sr} \leq x\}$. From the PDF of $g_{\rm br}$ and $g_{\rm sr}$, we can derive that
\begin{align}
F_{{\gamma}^{\rm m,II}}(x) &= \int_0^{\frac{x}{{{C_{{\rm{br}}}}}}} {\frac{{{y^{{N_{\rm{t}}}{\rm{ - 1}}}}}}{{\left( {{N_{\rm{t}}} - 1} \right)!}}{e^{ - y}}\int_0^{\frac{{x - {C_{{\rm{br}}}}y}}{{{C_{{\rm{sr}}}}}}} {{e^{ - z}}{\rm{d}}z} {\rm{d}}y}\nonumber \\
&=-e^{-\frac{x}{C_{\rm br}}}\int_0^{\frac{x}{{{C_{{\rm{br}}}}}}} {\frac{{{y^{{N_{\rm{t}}}{\rm{ - 1}}}}}}{{\left( {{N_{\rm{t}}} - 1} \right)!}}{e^{ - \left( {1 - {C_{{\rm{br}}}}/{C_{{\rm{sr}}}}} \right)y}}{\rm{d}}y} + \int_0^{\frac{x}{{{C_{{\rm{br}}}}}}} {\frac{{{y^{{N_{\rm{t}}}{\rm{ - 1}}}}}}{{\left( {{N_{\rm{t}}} - 1} \right)!}}{e^{ - y}}{\rm{d}}y}.\label{App:MCSNR}
\end{align}
Denote $B_1(x)= \int_0^{\frac{x}{{{C_{{\rm{br}}}}}}} {\frac{{{y^{{N_{\rm{t}}}{\rm{ - 1}}}}}}{{\left( {{N_{\rm{t}}} - 1} \right)!}}{e^{ - \left( {1 - {C_{{\rm{br}}}}/{C_{{\rm{sr}}}}} \right)y}}{\rm{d}}y}$ and $B_2(x) = \int_0^{\frac{x}{{{C_{{\rm{br}}}}}}} {\frac{{{y^{{N_{\rm{t}}}{\rm{ - 1}}}}}}{{\left( {{N_{\rm{t}}} - 1} \right)!}}{e^{ - y}}{\rm{d}}y}$. To apply \eqref{eq:AppCDF} in deriving the expression of $B_1(x)$, we denote $\tau =  \left( {1 - {C_{{\rm{br}}}}/{C_{{\rm{sr}}}}} \right)y$. Then,
\begin{align}
B_1(x) &= {\left( {\frac{{{C_{{\rm{sr}}}}}}{{{C_{{\rm{sr}}}} - {C_{{\rm{br}}}}}}} \right)^{{N_{\rm{t}}}}}\int_0^{\frac{{x\left( {{C_{{\rm{sr}}}} - {C_{{\rm{br}}}}} \right)}}{{{C_{{\rm{sr}}}}{C_{{\rm{br}}}}}}} {\frac{{{\tau ^{{N_{\rm{t}}}{\rm{ - 1}}}}}}{{\left( {{N_{\rm{t}}} - 1} \right)!}}{e^{ - \tau }}{\rm{d}}\tau }\nonumber\\
&=\left\{1-\exp\left[\frac{x(C_{\rm sr}-C_{\rm br})}{C_{\rm sr}C_{\rm br}}\right]\sum_{n=0}^{N_{\rm t}-1}{\frac{\left[\frac{x(C_{\rm sr}-C_{\rm br})}{C_{\rm sr}C_{\rm br}}\right]^n}{n!}}\right\}\left(\frac{C_{\rm sr}}{C_{\rm sr}-C_{\rm br}}\right)^{N_{\rm t}}\nonumber.
\end{align}
Similarly, we can derive that $B_2(x) = 1-\exp(-\frac{x}{C_{\rm br}}) \sum_{n=0}^{N_{\rm t}-1}{\frac{(\frac{x}{C_{\rm br}})^n}{n!}}$. Substituting $B_1(x)$ and $B_2(x)$ into \eqref{App:MCSNR}, we can obtain \eqref{eq:CDFmII}.
\end{proof}

\section{Proof of Property \ref{P:propertyRA}}
\label{App:Ploss}
\renewcommand{\theequation}{C.\arabic{equation}}
\setcounter{equation}{0}
\begin{proof}
To prove this property, we first show that the packet loss probability in \eqref{eq:loss} (i.e., $\Pr\{\mathcal{L}^{\rm m,II}|\mu_{\rm{sb}},\mu_{\rm{br}},\mu _{\rm{sr}}\} = \varepsilon^{\rm d}_{{\rm sr,}1}(1 - {\varepsilon^{\rm c, II} _{\rm{sb}}}){\varepsilon_2^{\rm{m,II}}} + \varepsilon^{\rm d}_{{\rm sr,}1} {\varepsilon^{\rm c, II} _{\rm{sb}}}\varepsilon^{\rm d}_{{\rm sr,}2}$) strictly increases with $r_{\rm A}^{\rm d}$.

The decoding error probability of cellular links does not depend on $r_{\rm A}^{\rm d}$, and hence ${\varepsilon^{\rm c, II} _{\rm{sb}}}$ is independent of $r_{\rm A}^{\rm d}$. From \eqref{eq:eu}, we can see that $\varepsilon^{\rm d}_{{\rm sr,}1}$ strictly decreases with $\mu _{\rm{sr}}$. Since $\alpha$ in  \eqref{eq:muHbr} is positive, $\mu _{\rm{sr}}$ strictly decreases with $r_{\rm A}^{\rm d}$. Thus, $\varepsilon^{\rm d}_{{\rm sr,}1}$ strictly increases with $r_{\rm A}^{\rm d}$. Similarly, we can prove that $\varepsilon^{\rm d}_{{\rm sr,}2}$ and ${\varepsilon_2^{\rm{m,II}}}$ also strictly increase with $r_{\rm A}^{\rm d}$. Therefore, the packet loss probability in \eqref{eq:loss} strictly increases with $r_{\rm A}^{\rm d}$.

Moreover, $\mu_{\rm c}$ and $\mu _{\rm{sr}}$ strictly increases with $\delta_{\rm c}$ and $\delta_{\rm sr}$, respectively, and hence \eqref{eq:loss} strictly decreases with $\delta_{\rm c}$ and $\delta_{\rm sr}$. Since $\Pr\{\Pr \{ {\mathcal{L}^{\rm{m,II}}}|{R_{\rm cell},r_{\rm A}^{\rm d},\delta_{\rm c}, \delta_{\rm sr}}\}\leq {\varepsilon}_{\max}\}$ equals to the left hand side of $\eqref{eq:PAHII}$, to prove Property \ref{P:propertyRA}, we only need to prove
\begin{align}\label{eq:B1}
\Pr\{\Pr \{ {\mathcal{L}^{\rm{m,II}}}|{R_{\rm cell},r_2,\delta_{\rm c}, \delta_{\rm sr}}\}\leq {\varepsilon}_{\max}\} < \Pr\{\Pr \{ {\mathcal{L}^{\rm{m,II}}}|{R_{\rm cell},r_1,\delta_{\rm c}, \delta_{\rm sr}}\}\leq {\varepsilon}_{\max}\}
\end{align}
for any two communication distances that satisfies $r_1 < r_2$. According to the definition of indicator function, \eqref{eq:B1} can be re-expressed as
\begin{align}\label{eq:B2}
{\mathbb{E}}\left\{{\mathbbm{1}}\left(\Pr \{ {\mathcal{L}^{\rm{m,II}}}|{R_{\rm cell},r_2,\delta_{\rm c}, \delta_{\rm sr}}\}\leq {\varepsilon}_{\max}\right)\right\} < {\mathbb{E}}\left\{{\mathbbm{1}}\left(\Pr \{ {\mathcal{L}^{\rm{m,II}}}|{R_{\rm cell},r_1,\delta_{\rm c}, \delta_{\rm sr}}\}\leq {\varepsilon}_{\max}\right)\right\}.
\end{align}
Since $\Pr \{ {\mathcal{L}^{\rm{m,II}}}|{R_{\rm cell},r_{\rm A}^{\rm d},\delta_{\rm c}, \delta_{\rm sr}}\}$ strictly increases with $r_{\rm A}^{\rm d}$ and strictly decreases with $\delta_{\rm c}$ and $\delta_{\rm sr}$, the $(\delta_{\rm c},\delta_{\rm sr})$-plane can be divided into the following three regions, denoted as ${\mathcal{S}}_1$, ${\mathcal{S}}_2$, and ${\mathcal{S}}_3$, respectively. For any $(\delta_{\rm c},\delta_{\rm sr}) \in {\mathcal{S}}_1$, $\Pr \{ {\mathcal{L}^{\rm{m,II}}}|{R_{\rm cell},r_1,\delta_{\rm c}, \delta_{\rm sr}}\}<\Pr \{ {\mathcal{L}^{\rm{m,II}}}|{R_{\rm cell},r_2,\delta_{\rm c}, \delta_{\rm sr}}\} \leq  {\varepsilon}_{\max}$. For any $(\delta_{\rm c},\delta_{\rm sr}) \in {\mathcal{S}}_2$, $\Pr \{ {\mathcal{L}^{\rm{m,II}}}|{R_{\rm cell},r_1,\delta_{\rm c}, \delta_{\rm sr}}\}\leq{\varepsilon}_{\max}<\Pr \{ {\mathcal{L}^{\rm{m,II}}}|{R_{\rm cell},r_2,\delta_{\rm c}, \delta_{\rm sr}}\}$. For any $(\delta_{\rm c},\delta_{\rm sr}) \in {\mathcal{S}}_3$, ${\varepsilon}_{\max} < \Pr \{ {\mathcal{L}^{\rm{m,II}}}|{R_{\rm cell},r_1,\delta_{\rm c}, \delta_{\rm sr}}\} < \Pr \{ {\mathcal{L}^{\rm{m,II}}}|{R_{\rm cell},r_2,\delta_{\rm c}, \delta_{\rm sr}}\}$.
Then, we have
\begin{align}
&{\mathbb{E}}\left\{{\mathbbm{1}}\left(\Pr \{ {\mathcal{L}^{\rm{m,II}}}|{R_{\rm cell},r_2,\delta_{\rm c}, \delta_{\rm sr}}\}\leq {\varepsilon}_{\max}\right)\right\} \nonumber\\
=& 1\times\Pr\{(\delta_{\rm c},\delta_{\rm sr}) \in {\mathcal{S}}_1\}+0\times\Pr\{(\delta_{\rm c},\delta_{\rm sr}) \in {\mathcal{S}}_2\}+0\times\Pr\{(\delta_{\rm c},\delta_{\rm sr}) \in {\mathcal{S}}_3\}\nonumber\\
=& \Pr\{(\delta_{\rm c},\delta_{\rm sr}) \in {\mathcal{S}}_1\}.\label{eq:Er2}
\end{align}

Similarly, we can derive that
\begin{align}
{\mathbb{E}}\left\{{\mathbbm{1}}\left(\Pr \{ {\mathcal{L}^{\rm{m,II}}}|{R_{\rm cell},r_1,\delta_{\rm c}, \delta_{\rm sr}}\}\leq {\varepsilon}_{\max}\right)\right\} = \Pr\{(\delta_{\rm c},\delta_{\rm sr}) \in {\mathcal{S}}_1\} + \Pr\{(\delta_{\rm c},\delta_{\rm sr}) \in {\mathcal{S}}_2\}.\label{eq:Er1}
\end{align}
From \eqref{eq:Er2} and \eqref{eq:Er1}, we can obtain \eqref{eq:B2}. This completes the proof.
\end{proof}

\section{Proof of Property \ref{P:propertySNR2}}
\label{App:mu2}
\renewcommand{\theequation}{D.\arabic{equation}}
\setcounter{equation}{0}
\begin{proof}
	We first consider two asymptotic cases: $\mu_{\rm br} = 0$ and $\mu_{\rm br} \to \infty$. When  $\mu_{\rm br} = 0$, $\gamma^{\rm m, I} = \frac{{{\mu _{{\rm{sr}}}}{g_{{\rm{sr}}}}P_{\rm{s}}^{\rm{t}}}}{{{N_0}W}}$, which is equal to the SNR with D2D model. When  $\mu_{\rm br}  \to \infty$, $\gamma^{\rm m, I} = \beta  = \frac{{{\mu _{{\rm{sb}}}}{g_{{\rm{sb}}}}P_{\rm{s}}^{\rm{t}}}}{{{N_0}W}}$, i.e., the UL SNR in the first phase. Moreover, from \eqref{eq:SNRHI}, we can derive that
	\begin{align}\label{eq:dSNR}
	\frac{{{\rm{d}}{\gamma ^{{\rm{m}},{\rm{I}}}}}}{{{\rm{d}}{\mu _{{\rm{br}}}}}} = \frac{{{g_{{\rm{br}}}}P_{\rm{b}}^{\rm{t}}(\beta  + 1){N_{\rm{t}}}\left( {\beta {N_0}W - {\mu _{{\rm{sr}}}}{g_{{\rm{sr}}}}P_{\rm{s}}^{\rm{t}}} \right)}}{{{{\left[ {{\mu _{{\rm{br}}}}{g_{{\rm{br}}}}P_{\rm{b}}^{\rm{t}} + (\beta  + 1){N_{\rm{t}}}{N_0}W} \right]}^2}}}.
	\end{align}
	If $\beta < \frac{{{\mu _{{\rm{sr}}}}{g_{{\rm{sr}}}}P_{\rm{s}}^{\rm{t}}}}{{{N_0}W}}$, then \eqref{eq:dSNR} is negative, and hence $\gamma^{\rm m, I}$ decreases with $\mu_{\rm br}$. Otherwise, $\gamma^{\rm m, I}$ increases with $\mu_{\rm br}$. Therefore, if $\beta < \frac{{{\mu _{{\rm{sr}}}}{g_{{\rm{sr}}}}P_{\rm{s}}^{\rm{t}}}}{{{N_0}W}}$, $\gamma^{\rm m, I}$ decreases from $\frac{{{\mu _{{\rm{sr}}}}{g_{{\rm{sr}}}}P_{\rm{s}}^{\rm{t}}}}{{{N_0}W}}$ to $\beta$ as $\mu_{\rm br}$ increases from $0$ to $\infty$. Otherwise, $\gamma^{\rm m, I}$ increases from $\frac{{{\mu _{{\rm{sr}}}}{g_{{\rm{sr}}}}P_{\rm{s}}^{\rm{t}}}}{{{N_0}W}}$ to $\beta$ as $\mu_{\rm br}$. The proof follows.
\end{proof}

\bibliographystyle{IEEEtran}
\bibliography{ref}

\end{document}